\documentclass[Journal,twocolumn]{IEEEtran}
\IEEEoverridecommandlockouts
\usepackage{color}
\usepackage{graphicx}
\usepackage{amsmath}
\usepackage{amssymb}
\usepackage{algorithm}
\usepackage{algorithmic}
\usepackage{amsmath}
\usepackage{multirow}
\usepackage{booktabs}
\usepackage{array}
\usepackage{amsthm}
\usepackage{stfloats}
\usepackage{caption}
\usepackage{subfigure}
\usepackage{bm}
\usepackage{booktabs}
\usepackage{setspace}
\usepackage{url}
\usepackage{arydshln}

\newcommand{\be}{\begin{equation}}
\newcommand{\ee}{\end{equation}}
\newcommand{\bea}{\begin{eqnarray}}
\newcommand{\eea}{\end{eqnarray}}
\newcommand{\ba}{\begin{array}}
\newcommand{\ea}{\end{array}}

\newcommand{\non}{\nonumber}

\renewcommand{\thesubsubsection}{\thesubsection.\arabic{subsubsection}}

\renewcommand{\baselinestretch}{1}
\allowdisplaybreaks[4]

\title{SNR/CRB-Constrained Joint Beamforming and Reflection Designs for RIS-ISAC Systems
\thanks{Manuscript received January 20, 2023; revised June 14, 2023 and September 25, 2023; accepted November 29, 2023. This work is supported in part by the National Natural Science Foundation of China (Grant No. 62371090 and 62071083), in part by Liaoning Applied Basic Research Program (Grant No. 2023JH2/101300201), and in part by Dalian Science and Technology Innovation Project (Grant No. 2022JJ12GX014). The work of A. Lee Swindlehurst is supported by the U.S. National Science Foundation Grants CCF-2225575 and CCF-2322191. The associate editor coordinating the review of this manuscript and approving it for publication is Prof. J. Yuan. \textit{(Corresponding author: Ming Li.)}}
\thanks{Part of this paper has been presented at the European Signal Processing Conference (EUSIPCO), 2022 \cite{Liu-EUSIPCO-22}.}
\thanks{R. Liu was with the School of Information and Communication Engineering, Dalian University of Technology, Dalian 116024, China, and now with the Center for Pervasive Communications and Computing, University of California, Irvine, CA 92697, USA (e-mail: rangl2@uci.edu). }
\thanks{M. Li is with the School of Information and Communication Engineering, Dalian University of Technology, Dalian 116024, China (e-mail: mli@dlut.edu.cn).}
\thanks{Q. Liu is with the School of Computer Science and Technology, Dalian University of Technology, Dalian 116024, China, and also with The Key Laboratory of Social Computing and Cognitive Intelligence, Ministry of Education (e-mail: qianliu@dlut.edu.cn).}
\thanks{A. L. Swindlehurst is with the Center for Pervasive Communications and Computing, University of California, Irvine, CA 92697, USA (e-mail: swindle@uci.edu).} }

\author{Rang Liu,~\IEEEmembership{Member,~IEEE,}
        Ming Li,~\IEEEmembership{Senior Member,~IEEE,}
        Qian Liu,~\IEEEmembership{Member,~IEEE,}\\
        and A. Lee Swindlehurst,~\IEEEmembership{Fellow,~IEEE}}

\begin{document}

\maketitle
\pagestyle{empty}
\thispagestyle{empty}

\begin{abstract}
In this paper, we investigate the integration of integrated sensing and communication (ISAC) and reconfigurable intelligent surfaces (RIS) for providing wide-coverage and ultra-reliable communication and high-accuracy sensing functions.
In particular, we consider an RIS-assisted ISAC system in which a multi-antenna base station (BS) simultaneously performs multi-user multi-input single-output (MU-MISO) communications and radar sensing with the assistance of an RIS.
We focus on both target detection and parameter estimation performance in terms of the signal-to-noise ratio (SNR) and Cram\'{e}r-Rao bound (CRB), respectively.
Two optimization problems are formulated for maximizing the achievable sum-rate of the multi-user communications under an SNR constraint for target detection or a CRB constraint for parameter estimation, the transmit power budget, and the unit-modulus constraint of the RIS reflection coefficients.
Efficient algorithms are developed to solve these two complicated non-convex problems. We then extend the proposed joint design algorithms to the scenario with imperfect self-interference cancellation. 
Extensive simulation results demonstrate the advantages of the proposed joint beamforming and reflection designs compared with other schemes.
In addition, it is shown that more RIS reflection elements bring larger performance gains for direct-of-arrival (DoA) estimation than for target detection.
\end{abstract}

\section{Introduction}

While wireless communication and radar sensing have been separately developed for decades, integrated sensing and communication (ISAC) has recently arising as a promising technology.
The integration of communication and sensing systems has evolved from coexistence, cooperation, to co-design, thanks to similar hardware platforms and signal processing algorithms and the same evolution direction towards high-frequency wideband multi-antenna systems.
ISAC not only allows communication and radar systems to share spectrum resources, but also enables a fully-shared platform transmitting unified waveforms to simultaneously perform communication and radar sensing functions, which significantly improves the spectral/energy/hardware efficiency  \cite{Zhang-ICST-2022}-\cite{Liu CST 2022}. 
It is foreseeable that in next-generation wireless networks, ISAC will be a supporting technology for diverse applications that require both high-quality ubiquitous wireless communications and high-accuracy sensing, e.g., smart home/factory, vehicular networks, etc.
Therefore, researchers from both academia and industry have explored various ISAC implementations \cite{Liu 2022}.

Advanced signal processing techniques have been investigated for designing the dual-functional transmit waveforms. 
Moreover, multi-input multi-output (MIMO) architectures are also widely considered in ISAC systems.
By exploiting the spatial degrees of freedom (DoFs) of MIMO architectures, the waveform diversity for radar sensing and the beamforming gains and spatial multiplexing for communications can be greatly improved  \cite{Liu-TSP-2020}.
Therefore, the transmit waveforms/beamforming design for ISAC systems is a key problem \cite{Zhang-JSTSP-2021}.
Extensive investigations have been conducted using various communication and radar sensing metrics \cite{Tong-JSTSP-2021}-\cite{Wen-TSP-2023}.
In addition, since radar sensing functions rely on analyzing received signal echoes, the receive beamforming is also jointly optimized to improve radar sensing performance \cite{Liu-JSAC-2022}.
Although these approaches greatly enhance communication and radar sensing functions, performance deterioration is still inevitable when encountering harsh propagation conditions.
In such complex electromagnetic environments, the use of recently emerged reconfigurable intelligent surface (RIS) technology \cite{Renzo-JSAC-2020}-\cite{Ayanoglu 2022} provides a potentially revolutionary solution.

RIS technology has been regarded as another key enabler for future wireless networks owing to its capability of efficiently and intelligently shaping the propagation environment.
An RIS is generally a two-dimensional meta-surface consisting of many passive reflecting elements that can be independently adjusted.
By controlling specific parameters of the electronic circuits associated with each element, electromagnetic characteristics of the incident signals can be tuned, e.g., amplitude, phase-shift, etc.
Passive beamforming gains can be achieved by cooperatively and intelligently adjusting these reflecting elements \cite{Huang-TWC-2019}-\cite{Liu TWC 2021}.
The deployment of RIS establishes non-line-of-sight (NLoS) links between transmitter and receivers, which expands coverage and introduces additional DoFs for improving system performance.
Therefore, the use of RIS in various wireless networks has been extensively investigated.

Inspired by extensive applications of RIS to various wireless communication systems \cite{Wu TCOM 2021}, researchers have begun exploring the deployment of RIS in radar sensing \cite{Song-arXiv-2022} and ISAC systems \cite{Liu-WC-2022}, \cite{Chepuri 2022}.
Studies for different RIS deployment scenarios with different communication and radar sensing metrics have been carried out \cite{Wang-TVT-2021a}-\cite{Liu-JSTSP-2022}.
The initial works \cite{Wang-TVT-2021a}-\cite{Luo TVT 2022} assumed that the RIS only assists with communications since it is deployed near the users with negligible impact on the target echoes.
To further exploit the benefits of RIS, especially in improving radar sensing performance, the authors in \cite{Song 2022}-\cite{Liao-TCOM-2023} assumed that the RIS is deployed where the direct links between the base station (BS) and the users/target are blocked.
However, it is likely more common that both the direct and reflected links contribute to communication and radar sensing functions, as considered in \cite{Yan JCS 2022} and \cite{Liu-JSTSP-2022}.
In particular, the authors in \cite{Yan JCS 2022} focused on the single-user scenario and used the signal-to-noise ratio (SNR) metric for both communication and target detection, while the authors in \cite{Liu-JSTSP-2022} studied the more common multi-user scenario in the presence of clutter and proposed to maximize the radar signal-to-interference-plus-noise ratio (SINR) under communication constraints.
However, the non-linear spatial-temporal beamforming assumed in \cite{Liu-JSTSP-2022} requires more complicated hardware architectures and more complex algorithms. 
Moreover, in addition to the target detection function, parameter estimation is also an important task in radar sensing and should be further explored.

Motivated by the above discussion, in this paper we focus on a general RIS-assisted ISAC system and consider both target detection and parameter estimation functions for the sensing component.
In particular, we consider an RIS-assisted ISAC system in which a multi-antenna BS delivers data to multiple single-antenna users and simultaneously senses one point-like target with the assistance of an RIS.
Our goal is to jointly design the BS transmit/receive beamforming and RIS reflection coefficients to maximize multi-user communication performance as well as guarantee target detection or direct-of-arrival (DoA) estimation performance.
The main contributions of this paper are summarized as follows.
\begin{itemize}
  \item First, we model the signals received at the users and the BS receive array, and then derive performance metrics for communications and radar sensing. Specifically, the sum-rate metric is utilized to evaluate the multi-user communications performance. For target detection, we show that the radar output SNR is positively proportional to the detection probability with a fixed probability of false alarm and derive a worst-case radar SNR as the performance metric. For target DoA estimation in the considered general case, we derive the Cram\'{e}r-Rao bound (CRB) for estimating the target DoAs with respect to the BS and the RIS, which to our knowledge has not been investigated previously in the literature. 
  \item Next, we formulate two optimization problems that maximize the sum-rate for multi-user communications subject to a radar SNR constraint for target detection or a CRB constraint for DoA estimation, the transmit power budget, and the unit-modulus constraint of the RIS reflection coefficients. In order to efficiently solve these two complicated non-convex problems, we develop two algorithms based on fractional programming (FP), majorization-minimization (MM), alternative direction method of multipliers (ADMM), and some sophisticated transformations to jointly solve for the BS transmit/receive beamforming and RIS reflection coefficients.
  \item Finally, we provide extensive simulation studies to verify the effectiveness of the proposed schemes and associated algorithms. It is shown that the proposed designs achieve notably higher sum-rates compared with other schemes. We also present the enhanced beampattern to visually demonstrate the dual communications and radar sensing functions of the considered RIS-assisted ISAC system.

\end{itemize}

\textit{Notation}: Boldface lower-case and upper-case letters indicate column vectors and matrices, respectively.
$(\cdot)^*$, $(\cdot)^T$, $(\cdot)^H$, and $(\cdot)^{-1}$  denote the conjugate, transpose, transpose-conjugate, and inverse operations, respectively.
$\mathbf{I}_M$ indicates an $M\times M$ identity matrix.
$\mathbb{C}$ and $\mathbb{R}$ denote the sets of complex numbers and real numbers, respectively.
$| a |$, $\| \mathbf{a} \|$, and $\| \mathbf{A} \|_F$ are the magnitude of a scalar $a$, the norm of a vector $\mathbf{a}$, and the Frobenius norm of a matrix $\mathbf{A}$, respectively.
$\mathbb{E}\{\cdot\}$ represents statistical expectation.
$\text{Tr}\{\mathbf{A}\}$ takes the trace of the matrix $\mathbf{A}$ and $\text{vec}\{\mathbf{A}\}$ vectorizes the matrix $\mathbf{A}$.
$\otimes$ denotes the Kronecker product.
$\mathfrak{R}\{\cdot\}$ and $\mathfrak{I}\{\cdot\}$ denote the real and imaginary parts of a complex number, respectively.
$\angle{a}$ is the angle of complex-valued $a$.

\begin{figure}[t]
  \centering
  \includegraphics[width=\linewidth]{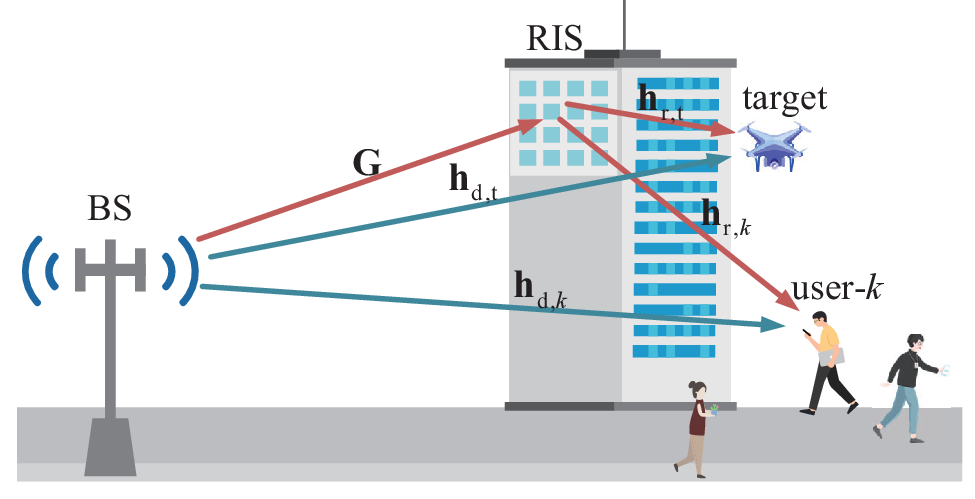}\\
  \caption{An RIS-assisted ISAC system.}\label{fig:system model}\vspace{-0.2 cm}
\end{figure}

\section{System and Signal Models}\label{sec:system and signal models}

We consider an RIS-assisted ISAC system as shown in Fig. \ref{fig:system model}, where a colocated multi-antenna BS performs multi-user communications and radar sensing with the assistance of an $N$-element RIS.
In particular, the BS, which is equipped with $M$ transmit antennas and $M$ receive antennas arranged as uniform linear arrays (ULAs) with half-wavelength spacing, simultaneously serves $K$ single-antenna users and senses one target.
With the aid of advanced self-interference mitigation techniques \cite{Sabharwal-JSAC-2014}, we assume that the BS operates in full-duplex mode with perfect self-interference mitigation.
In this work, we focus on two common and crucial radar sensing tasks: \textit{target detection}, which aims to identify whether a target of interest is present, and \textit{parameter estimation}, where the DoAs of the target signals with respect to the BS and the RIS are determined.

In order to simultaneously enable satisfactory communication and radar sensing functions, the dual-functional signal that is transmitted in the $l$-th time slot is given by \cite{Liu-TSP-2020}
\be
\mathbf{x}[l] = \mathbf{W}_\text{c}\mathbf{s}_\text{c}[l] + \mathbf{W}_\text{r}\mathbf{s}_\text{r}[l]  = \mathbf{W}\mathbf{s}[l],
\ee
where $\mathbf{s}_\text{c}[l] \in\mathbb{C}^K$ contains the communication symbols for the $K$ users with $\mathbb{E}\{\mathbf{s}_\text{c}[l]\mathbf{s}_\text{c}^H[l]\}=\mathbf{I}_K$, $\mathbf{s}_\text{r}[l] \in\mathbb{C}^M$ includes $M$ individual radar waveforms with $\mathbb{E}\{\mathbf{s}_\text{r}[l]\mathbf{s}_\text{r}^H[l]\}=\mathbf{I}_M$, $\mathbb{E}\{\mathbf{s}_\text{c}[l]\mathbf{s}_\text{r}^H[l]\}=\mathbf{0}$, and  $\mathbf{W}_\text{c}\in\mathbb{C}^{M\times K}$ and $\mathbf{W}_\text{r}\in\mathbb{C}^{M\times M}$ denote the beamforming matrices for the  communication symbols and radar waveforms, respectively.
In addition, we define the combined beamforming matrix $\mathbf{W} \triangleq [\mathbf{W}_\text{c}~\mathbf{W}_\text{r}]$ and symbol vector $\mathbf{s}[l]  \triangleq [\mathbf{s}_\text{c}^T[l] ~\mathbf{s}_\text{r}^T[l] ]^T$ for brevity.
Then, the compound received signal at the $k$-th user can be expressed as
\be\label{eq:yk}
y_k[l]  = (\mathbf{h}_{\text{d},k}^T + \mathbf{h}_{\text{r},k}^T\bm{\Phi}\mathbf{G})\mathbf{x}[l] + n_k[l],
\ee
where $\mathbf{h}_{\text{d},k}\in\mathbb{C}^M$, $\mathbf{h}_{\text{r},k}\in\mathbb{C}^N$, and $\mathbf{G}\in\mathbb{C}^{N\times M}$ denote the channels between the BS/RIS and the $k$-th user, and between the BS and the RIS, respectively.
The channels $\mathbf{h}_{\text{d},k}$, $\mathbf{h}_{\text{r},k},~\forall k$, and $\mathbf{G}$ are assumed to follow Rician fading, e.g., the channel $\mathbf{G}$ is formulated as $\mathbf{G} = \alpha_\text{G}\sqrt{\frac{\kappa}{\kappa+1}}\mathbf{G}_\text{LoS} + \alpha_\text{G} \sqrt{\frac{1}{\kappa+1}}\mathbf{G}_\text{NLoS}$, where $\alpha_\text{G}$ is the distance dependent pathloss, $\kappa$ represents the Rician factor, $\mathbf{G}_\text{LoS}$ is the LoS component that depends on the problem geometry, and $\mathbf{G}_\text{NLoS}$ denotes the NLoS Rayleigh fading components with zero mean and unit variance.
In particular, $\mathbf{G}_\text{LoS} = \mathbf{a}_N^T(\theta_\text{RB})\mathbf{a}_M(\theta_\text{BR})$, where the steering vector is $\mathbf{a}_N(\theta_\text{RB})\triangleq[1,e^{-\jmath\pi\sin\theta_\text{RB}},\ldots, e^{-\jmath(N-1)\pi\sin\theta_\text{RB}}]^T$, and $\theta_\text{RB}$ and $\theta_\text{BR}$ represent the DoA and direct-of-departure (DoD).
We assume that the overall channel is known at the BS given advanced channel estimation approaches for RIS-aided communication systems \cite{Hu-TCOM-2021}-\cite{Swindle-Proc-2022}.
The RIS reflection matrix is defined as $\bm{\Phi}\triangleq\text{diag}\{\bm{\phi}\}$, where $\bm{\phi}\triangleq[\phi_1,\phi_2,\ldots,\phi_N]^T$ is the vector of reflection coefficients satisfying $|\phi_n|=1,~\forall n$.
The scalar term $n_k[l]\sim\mathcal{CN}(0,\sigma_k^2)$ is additive white Gaussian noise (AWGN) at the $k$-th user.

The transmitted signal will reach the target via both the direct and reflected links and then also be reflected back to the BS through these two links.
It is noted that the RIS naturally operates in full-duplex mode and is free of self-interference since it is composed of passive reflecting elements \cite{Wu TCOM 2021}.
Thus,
the baseband echo signal which is reflected by the target and then collected by the BS receive array can be expressed as
\be
\mathbf{y}_\text{r}[l] = \alpha_\text{t}(\mathbf{h}_{\text{d},\text{t}} + \mathbf{G}^T\bm{\Phi}\mathbf{h}_{\text{r},\text{t}})(\mathbf{h}^T_{\text{d},\text{t}} + \mathbf{h}^T_{\text{r},\text{t}}\bm{\Phi}\mathbf{G})\mathbf{W}\mathbf{s}[l] + \mathbf{n}_\text{r}[l],
\ee
where $\alpha_\text{t}\sim\mathcal{CN}(0,\sigma_\text{t}^2)$ is the radar cross section (RCS), $\mathbf{h}_{\text{d},\text{t}}\in\mathbb{C}^M$ and $\mathbf{h}_{\text{r},\text{t}}\in\mathbb{C}^N$ respectively represent the channels between the BS/RIS and the target, and $\mathbf{n}_\text{r}[l]\sim\mathcal{CN}(\mathbf{0},\sigma_\text{r}^2\mathbf{I}_M)$ is the AWGN.
As typically done in radar sensing, we will assume that the BS/RIS-target links are LoS.
Specifically, $\mathbf{h}_{\text{d},\text{t}} = \alpha_\text{dt}\mathbf{a}_M(\theta_1)$ and $\mathbf{h}_{\text{r},\text{t}}=\alpha_\text{rt}\mathbf{a}_N(\theta_2)$, where $\theta_1$ and $\theta_2$ are the DoAs of the target with respect to the BS and the RIS, respectively.
The DoAs can be obtained by using the classic localization methods \cite{Bekkerman TSP 2006} or maximum likelihood estimation algorithms based on previous observations. With the knowledge of $\theta_1$ and $\theta_2$, the target detection and parameter estimation performance can be further improved through beamforming designs at the BS.
In addition, thanks to the semi-static characteristic of the BS-RIS channel, a two-timescale channel estimation framework \cite{Hu-TCOM-2021} can be utilized to estimate $\mathbf{G}$.
Radar sensing relies on analyzing received echo signals over $L$ samples, which we combine together and denote as
\be\label{eq:received echo signal Yr}
\mathbf{Y}_\text{r}  = \alpha_\text{t}\mathbf{H}_\text{t}(\bm{\phi})\mathbf{W}\mathbf{S} + \mathbf{N}_\text{r},
\ee
where we define $\mathbf{H}_\text{t}(\bm{\phi})\triangleq(\mathbf{h}_{\text{d},\text{t}} + \mathbf{G}^T\bm{\Phi}\mathbf{h}_{\text{r},\text{t}})(\mathbf{h}^T_{\text{d},\text{t}} + \mathbf{h}^T_{\text{r},\text{t}}\bm{\Phi}\mathbf{G})$, and the symbol/noise matrices as $\mathbf{S} \triangleq [\mathbf{s}[1],\mathbf{s}[2],\ldots,\mathbf{s}[L]]$ and $\mathbf{N}_\text{r} \triangleq [\mathbf{n}_\text{r}[1],\mathbf{n}_\text{r}[2],\ldots,\mathbf{n}_\text{r}[L]]$, respectively.
It is noted that the equivalent channel $\mathbf{H}_\text{t}(\bm{\phi})$ is known at the BS based on the above assumptions.

\section{Performance Metrics}\label{sec:performance metrics}

In this section, we separately derive the performance metrics for communications and radar sensing, which are crucial in formulating the optimization problems for the considered system.
The typical sum-rate metric is formulated to evaluate the performance of multi-user communications.
For radar sensing, we consider both target detection and parameter estimation performance in terms of SNR and CRB, respectively.

\subsection{Sum-rate for Multi-user Communications}

The achievable sum-rate is the most widely used metric to evaluate the performance of multi-user communications.
Based on the signal model in (\ref{eq:yk}), the SINR of the $k$-th user is
\be
\text{SINR}_k = \frac{|\mathbf{h}^T_k(\bm{\phi})\mathbf{w}_k|^2}
{\sum_{j\neq k}^{K+M}|\mathbf{h}^T_k(\bm{\phi})\mathbf{w}_j|^2+\sigma_k^2},
\ee
where for conciseness we define $\mathbf{h}_k(\bm{\phi}) \triangleq \mathbf{h}_{\text{d},k} + \mathbf{G}^T\bm{\Phi}\mathbf{h}_{\text{r},k}$ as the composite channel between the BS and the $k$-th user, and $\mathbf{w}_j$ as the $j$-th column of $\mathbf{W}$, i.e., $\mathbf{W} = [\mathbf{w}_1,\mathbf{w}_2,\ldots,\mathbf{w}_{K+M}]$.
The achievable sum-rate of the $K$ users is then given by $R = \sum_{k=1}^K\log_2(1+\text{SINR}_k)$.

\subsection{SNR for Target Detection in Sensing}

Target detection is a primary task in radar sensing.
In order to achieve better target detection performance, i.e., a higher probability of detection, the received echo signals are processed by a matched filter using the information of transmitted symbols $\mathbf{S}$ to improve the output SNR.
The received signals $\mathbf{Y}_\text{r}$ after the matched-filtering can be written as
\be\label{eq:Yr after match filter}
\widetilde{\mathbf{Y}}_\text{r} = \alpha_\text{t}\mathbf{H}_\text{t}(\bm{\phi})\mathbf{WSS}^H + \mathbf{N}_\text{r}\mathbf{S}^H.
\ee
By defining $\widetilde{\mathbf{y}}_\text{r} \triangleq \text{vec}\{\widetilde{\mathbf{Y}}_\text{r}\}$, $\mathbf{w} \triangleq \text{vec}\{\mathbf{W}\}$, and $\widetilde{\mathbf{n}}_\text{r} \triangleq \text{vec}\{\mathbf{N}_\text{r}\mathbf{S}^H\}$, the vectorized signal can be expressed as
\be\label{eq:vec Yr after match}
\widetilde{\mathbf{y}}_\text{r}  = \alpha_\text{t}(\mathbf{SS}^H\otimes\mathbf{H}_\text{t}(\bm{\phi}))\mathbf{w}+ \widetilde{\mathbf{n}}_\text{r}.
\ee
Then, a receive filter/beamformer $\mathbf{u}\in\mathbb{C}^{M\times (K+M)}$ is applied to process $\widetilde{\mathbf{y}}_\text{r}$ and yields
\be\label{eq:target detection output}
\mathbf{u}^H\widetilde{\mathbf{y}}_\text{r}  = \alpha_\text{t}\mathbf{u}^H(\mathbf{SS}^H\otimes\mathbf{H}_\text{t}(\bm{\phi}))\mathbf{w} + \mathbf{u}^H\widetilde{\mathbf{n}}_\text{r}.
\ee

Thus, the hypothesis testing problem for the output of the radar receiver is expressed as
\be\label{eq:two hypo}
y = \bigg\{ \begin{array}{l l }
\mathcal{H}_0: \mathbf{u}^H\widetilde{\mathbf{n}}_\text{r},\\
\mathcal{H}_1: \alpha_\text{t}\mathbf{u}^H(\mathbf{SS}^H\otimes\mathbf{H}_\text{t}(\bm{\phi}))\mathbf{w} + \mathbf{u}^H\widetilde{\mathbf{n}}_\text{r}.\end{array} 
\ee
Noting that $\alpha_\text{t}\sim\mathcal{CN}(0,\sigma_\text{t}^2)$ and $\mathbf{H}_\text{t}(\bm{\phi})$ is known at the BS, we have the conditional probability distributions $y|\mathcal{H}_0\sim\mathcal{CN}(0,\eta_0)$ and $y|\mathcal{H}_1\sim\mathcal{CN}(0,\eta_1)$ with $\eta_0 = L\sigma_\text{r}^2\mathbf{u}^H\mathbf{u}$ and $\eta_1 = \sigma_\text{t}^2\mathbb{E}\big\{|\mathbf{u}^H(\mathbf{SS}^H\otimes\mathbf{H}_\text{t}(\bm{\phi}))\mathbf{w}|^2\big\}+L\sigma_\text{r}^2\mathbf{u}^H\mathbf{u}$.
The Neyman-Pearson detector to identify whether the target is present can be formulated as \cite{detection theory}, \cite{Bekkerman TSP 2006}: $T = |y|^2\underset{\mathcal{H}_0}{\overset{\mathcal{H}_1}{\gtrless}}\delta$, where the decision threshold $\delta$ is determined by the probability of false alarm $P_\text{FA}$.
Accordingly, the statistic distribution of $T$ is given by $T \sim \big\{ \begin{array}{r r }
\eta_0\chi_2^2,  \quad\mathcal{H}_0,\\
\eta_1\chi_2^2,  \quad\mathcal{H}_1,\end{array} $, where $\chi_2^2$ indicates the central chi-squared distribution with two DoFs.

In the sequel, the probabilities of detection $P_\text{D}$ and false alarm $P_\text{FA}$ can be calculated as
\begin{subequations}\label{eq:probs of detection and false alarm}
\begin{align}
P_\text{FA}&= \text{Pr}(T>\delta|\mathcal{H}_0) = 1-\mathfrak{F}_{\chi_2^2}(\delta/\eta_0),\\
P_\text{D}&= \text{Pr}(T>\delta|\mathcal{H}_1) = 1-\mathfrak{F}_{\chi_2^2}(\delta/\eta_1),
\end{align}\end{subequations}
where $\text{Pr}(\cdot)$ denotes the probability and $\mathfrak{F}_{\chi_2^2}(\cdot)$ represents the central chi-squared distribution function with two DoFs.
For a desired $P_\text{FA}$, the achieved $P_\text{D}$ can then be calculated as
\be P_\text{D} = 1-\mathfrak{F}_{\chi_2^2}\big(\eta_0/\eta_1\mathfrak{F}_{\chi_2^2}^{-1}(1-P_\text{FA})\big),\ee
where $\mathfrak{F}_{\chi_2^2}^{-1}$ denotes the inverse central chi-squared distribution function with two DoFs.
Therefore, we have
\be
P_\text{D}\varpropto \eta_1/\eta_0 = \frac{\sigma_\text{t}^2\mathbb{E}\big\{|\mathbf{u}^H(\mathbf{SS}^H\otimes\mathbf{H}_\text{t}(\bm{\phi}))\mathbf{w}|^2\big\}}
{L\sigma_\text{r}^2\mathbf{u}^H\mathbf{u}} + 1.
\ee
We observe that $P_\text{D}$ is positively proportional to the radar SNR, which is expressed as
\be\label{eq:SNRt org}
\text{SNR}_\text{t} = \frac{\sigma_\text{t}^2\mathbb{E}\big\{|\mathbf{u}^H(\mathbf{SS}^H\otimes\mathbf{H}_\text{t}(\bm{\phi}))\mathbf{w}|^2\big\}}
{L\sigma_\text{r}^2\mathbf{u}^H\mathbf{u}}.
\ee
Thus, we use radar SNR to evaluate the target detection performance.
Considering that the numerator in (\ref{eq:SNRt org}) is complicated and difficult for optimization, we propose to optimize a lower bound of it.
In particular, since $\text{SNR}_\text{t}$ is a convex function of $\mathbf{SS}^H$ and $\mathbb{E}\{\mathbf{SS}^H\} = L\mathbf{I}_{K+M}$, the Jensen's inequality $\mathbb{E}\{f(x)\} \geq f(\mathbb{E}\{x\})$ is utilized to obtain a lower bound for the SNR as 
\be\label{eq:SNRt}
\text{SNR}_\text{t} \geq \frac{L\sigma_\text{t}^2|\mathbf{u}^H(\mathbf{I}_{K+M}\otimes\mathbf{H}_\text{t}(\bm{\phi}))\mathbf{w}|^2}
{\sigma_\text{r}^2\mathbf{u}^H\mathbf{u}},
\ee
which represents the worst-case achieved radar SNR.

\subsection{CRB for Parameter Estimation in Sensing}

Parameter estimation is another important task in radar sensing.
The accuracy of parameter estimation is usually measured by the Cram\'{e}r-Rao bound, which is a lower bound for any unbiased estimator.
In our considered setting, we focus on DoA estimation of $\bm{\theta}\triangleq[\theta_1, \theta_2]^T$.
To derive the CRB for estimating $\bm{\theta}$, we first vectorize the received signal $\mathbf{Y}_\text{r}$ as
\be\label{eq:vec Yr}
\mathbf{y}_\text{r}  = \alpha_\text{t}\text{vec}\{\mathbf{H}_\text{t}(\bm{\phi})\mathbf{WS}\}+ \mathbf{n}_\text{r},
\ee
where $\mathbf{n}_\text{r}\triangleq\text{vec}\{\mathbf{N}_\text{r}\}$.
We define the unknown target parameters as $\bm{\xi} \triangleq [\bm{\theta}^T, \bm{\alpha}^T]^T$ with $\bm{\alpha}\triangleq[\Re\{\alpha_\text{t}\}, \Im\{\alpha_\text{t}\}]^T$ and the noise-free echo signal as  $\bm{\eta}\triangleq\alpha_\text{t}\text{vec}\{\mathbf{H}_\text{t}(\bm{\phi})\}\mathbf{WS}$.

As presented in \cite{estimation theory}, for the complex observation $\mathbf{y}_\text{r} \sim \mathcal{CN}(\bm{\eta},\mathbf{R}_\text{r})$, $\mathbf{R}_\text{r} = \sigma_\text{r}^2\mathbf{I}_{ML}$, the $(i,j)$-th element of the Fisher information matrix (FIM) $\mathbf{F}_\text{IM}\in\mathbb{C}^{4\times 4}$ can be obtained as
\be
\mathbf{F}_\text{IM}(i,j) = \frac{2}{\sigma_\text{r}^2}\Re\Big\{\frac{\partial^H\bm{\eta}}
{\partial\xi_i}\frac{\partial\bm{\eta}}
{\partial\xi_j}\Big\}.
\label{eq:define FIM}\ee
In addition, the CRB matrix $\mathbf{C}$ is the inverse of $\mathbf{F}_\text{IM}$ and the diagonal elements of $\mathbf{C}$ represent the CRB for $\bm{\xi}$. In order to obtain a closed-form expression of the CRB for DoA estimation, we partition $\mathbf{F}_\text{IM}$ and $\mathbf{C}$ into $2\times 2$ blocks as
\be\label{eq:FIM partition}
\mathbf{F}_\text{IM} = \left[\!\begin{array}{cc}\mathbf{F}_{\bm{\theta}\bm{\theta}^T} & \mathbf{F}_{\bm{\theta}\bm{\alpha}^T}\\ \mathbf{F}^T_{\bm{\theta}\bm{\alpha}^T} & \mathbf{F}_{\bm{\alpha}\bm{\alpha}^T}\end{array}\!\right] = \left[\!\begin{array}{cc}\mathbf{C}_{\bm{\theta}\bm{\theta}^T} & \mathbf{C}_{\bm{\theta}\bm{\alpha}^T}\\ \mathbf{C}_{\bm{\alpha}\bm{\theta}^T} & \mathbf{C}_{\bm{\alpha}\bm{\alpha}^T}\end{array}\!\right]^{-1} = \mathbf{C}^{-1},
\ee
in which the expressions for the sub-matrices of $\mathbf{F}_\text{IM}$ can be derived according to (\ref{eq:vec Yr}) and (\ref{eq:define FIM}) as presented in Appendix A.
We should emphasize here that each element of $\mathbf{F}_\text{IM}$ is a function of both $\mathbf{W}$ and $\bm{\phi}$.
The CRB for estimating $\bm{\theta}$ can then be obtained as \cite{Bekkerman TSP 2006}
\begin{subequations}\label{eq:form CRB}\begin{align}\text{CRB}_{\theta_1}+\text{CRB}_{\theta_2}&= \text{Tr}\{\mathbf{C}_{\bm{\theta}\bm{\theta}^T}\}\\
&= \text{Tr}\{(\mathbf{F}_{\bm{\theta}\bm{\theta}^T}-\mathbf{F}_{\bm{\theta}\bm{\alpha}^T}
\mathbf{F}_{\bm{\alpha}\bm{\alpha}^T}^{-1}\mathbf{F}^T_{\bm{\theta}\bm{\alpha}^T})^{-1}\}.
\end{align}
\end{subequations}

\section{SNR-Constrained Joint Beamforming and Reflection Design}\label{sec:SNR constrained design}

In this section, we consider the SNR-constrained joint beamforming and reflection design problem.
We aim to jointly optimize the transmit beamforming matrix $\mathbf{W}$, the receive filter $\mathbf{u}$, and the reflection coefficients $\bm{\phi}$ to maximize the sum-rate, and satisfy the worst-case radar SNR requirement $\Gamma_\text{t}$, the transmit power budget $P_\text{t}$, and the unit-modulus property of the RIS reflecting coefficients.
Therefore, the optimization problem is formulated as
\begin{subequations}\label{eq:original problem}\begin{align}
&\underset{\mathbf{W},\mathbf{u},\bm{\phi}}\max~~\sum_{k=1}^K\log_2(1+\text{SINR}_k)\\
&\quad\text{s.t.}\quad~\frac{L\sigma_\text{t}^2|\mathbf{u}^H(\mathbf{I}_{K+M}\otimes\mathbf{H}_\text{t}(\bm{\phi}))\mathbf{w}|^2}
{\sigma_\text{r}^2\mathbf{u}^H\mathbf{u}} \geq \Gamma_\text{t}, \\
&\quad\quad\quad~~\|\mathbf{W}\|_F^2 \leq P_\text{t},\\
&\quad\quad\quad~~|\phi_n| = 1,~~\forall n.
\end{align}
\end{subequations}
It is obvious that the non-convex problem (\ref{eq:original problem}) is very difficult to solve due to the complicated objective function (\ref{eq:original problem}a) with $\log(\cdot)$ and fractional terms, the coupled variables in both the objective function (\ref{eq:original problem}a) and the radar SNR constraint (\ref{eq:original problem}b), and the unit-modulus constraint (\ref{eq:original problem}d).
In order to tackle these difficulties, we propose to utilize FP, MM, and ADMM methods to convert problem (\ref{eq:original problem}) into several tractable sub-problems and iteratively solve them.

\subsection{FP-based Transformation}

We start by converting the objective function (\ref{eq:original problem}a) into a more favorable polynomial expression based on FP.
As derived in \cite{Shen-TSP-2018}, by employing the Lagrangian dual reformulation and introducing an auxiliary variable $\mathbf{r}\triangleq[r_1,r_2,\ldots,r_K]^T$, the objective (\ref{eq:original problem}a) can be transformed to\setcounter{equation}{20}
\be\label{eq:after FP1}
\sum_{k=1}^K\!\log_2(1+r_k) - \sum_{k=1}^K\!r_k + \sum_{k=1}^K\!\frac{(1+r_k)|\mathbf{h}^T_k(\bm{\phi})\mathbf{w}_k|^2}
{\sum_{j=1}^{K+M}\!|\mathbf{h}^T_k(\bm{\phi})\mathbf{w}_j|^2\!+\!\sigma_k^2},
\ee
in which the variables $\mathbf{w}$ and $\bm{\phi}$ are taken out of the $\log(\cdot)$ function and coupled in the third fractional term.
Then, expanding the quadratic terms and introducing an auxiliary variable $\mathbf{c}\triangleq[c_1,c_2,\ldots,c_K]^T$, the expression in (\ref{eq:after FP1}) can be further converted to
\be\label{eq:new obj}\begin{aligned}
&\mathcal{F}(\mathbf{w},\bm{\phi},\mathbf{r},\mathbf{c}) \triangleq  \sum_{k=1}^K\log_2(1+r_k) - \sum_{k=1}^K r_k -\sum_{k=1}^K|c_k|^2\sigma_k^2 \\
& + \sum_{k=1}^K\!2\sqrt{1\!+\!r_k}\Re\{c_k^*\mathbf{h}^T_k(\bm{\phi})\mathbf{w}_k\} -\!\sum_{k=1}^K\!|c_k|^2\!\!\sum_{j=1}^{K+M}|\mathbf{h}^T_k(\bm{\phi})\mathbf{w}_j|^2.
\end{aligned}\ee

\newcounter{TempEqCnt}
\setcounter{TempEqCnt}{\value{equation}}
\setcounter{equation}{19}
\begin{figure*}[!t]
\begin{subequations}\label{eq:reformulate1}
\begin{align}
&(\mathbf{I}\otimes\mathbf{H}_\text{t}(\bm{\phi}))\mathbf{w} \non \\
& = \big(\mathbf{I}\!\otimes\!\mathbf{h}_{\text{d},\text{t}}\mathbf{h}_{\text{d},\text{t}}^T\big)\mathbf{w} + \text{vec}\big\{\mathbf{G}^T\text{diag}\{\mathbf{h}_{\text{r},\text{t}}\}\bm{\phi}\mathbf{h}_{\text{d},\text{t}}^T\mathbf{W}
+ \mathbf{h}_{\text{d},\text{t}}\bm{\phi}^T\text{diag}\{\mathbf{h}_{\text{r},\text{t}}\}\mathbf{G}\mathbf{W}
+ \mathbf{G}^T\text{diag}\{\mathbf{h}_{\text{r},\text{t}}\}\bm{\phi}\bm{\phi}^T\text{diag}\{\mathbf{h}_{\text{r},\text{t}}\}\mathbf{G}\mathbf{W}\big\}
\\& = \big(\mathbf{I}\!\otimes\!\mathbf{h}_{\text{d},\text{t}}\mathbf{h}_{\text{d},\text{t}}^T\big)\mathbf{w} +
\big(\!\underbrace{\mathbf{W}^T\mathbf{h}_{\text{d},\text{t}}\!\otimes\!\mathbf{G}^T\!\text{diag}\{\mathbf{h}_{\text{r},\text{t}}\}
\!+\!\mathbf{W}^T\mathbf{G}^T\!\text{diag}\{\mathbf{h}_{\text{r},\text{t}}\}\!\otimes\!\mathbf{h}_{\text{d},\text{t}}}_{\mathbf{F}}\!\big)\bm{\phi}
+ \big(\!\underbrace{\mathbf{W}^T\mathbf{G}^T\!\text{diag}\{\mathbf{h}_{\text{r},\text{t}}\}\!\otimes\!\mathbf{G}^T\!\text{diag}\{\mathbf{h}_{\text{r},\text{t}}\}}_{\mathbf{L}}\!\big)\text{vec}\{\bm{\phi}\bm{\phi}^T\}.
\label{eq:reformulate1b}
\end{align}\end{subequations}

\rule[-0pt]{18.5 cm}{0.05em}
\end{figure*}
\setcounter{equation}{\value{TempEqCnt}}

To facilitate the algorithm development, we attempt to re-arrange the new objective function (\ref{eq:new obj}) into explicit and compact forms with respect to $\mathbf{w}$ and $\bm{\phi}$, respectively.
By stacking the vectors $\mathbf{w}_j,~\forall j$, into $\mathbf{w}$ and applying $\mathbf{h}^T_k(\bm{\phi})\mathbf{w}_j = \mathbf{h}_{\text{d},k}^T\mathbf{w}_j + \mathbf{h}_{\text{r},k}^T\text{diag}\{\mathbf{G}\mathbf{w}_j\}\bm{\phi}$, the following  equivalent expressions for $\mathcal{F}(\mathbf{w},\bm{\phi},\mathbf{r},\mathbf{c})$ can be obtained:
\begin{subequations}\label{eq: new obj func}\begin{align}
\mathcal{F}(\mathbf{w},\bm{\phi},\mathbf{r},\mathbf{c}) &= \Re\{\mathbf{a}^H\mathbf{w}\}-\|\mathbf{B}\mathbf{w}\|^2 + \varepsilon_1\\
& = \Re\{\mathbf{g}^H\bm{\phi}\} - \bm{\phi}^H\mathbf{D}\bm{\phi} + \varepsilon_2,
\end{align}\end{subequations}
where the definitions of $\mathbf{a}$, $\mathbf{B}$, $\varepsilon_1$, $\mathbf{g}$, $\mathbf{D}$, and $\varepsilon_2$ can be easily obtained through basic matrix operations and are omitted here due to space limitations.
Now, we can clearly see that the re-formulated objective $\mathcal{F}(\mathbf{w},\bm{\phi},\mathbf{r},\mathbf{c})$ is a conditionally concave function with respect to each variable given the others, which allows us to iteratively solve for each variable as shown below.

\subsection{Block Update}

\subsubsection{Update $\mathbf{r}$ and $\mathbf{c}$}

Given the other variables, the optimization for the auxiliary variable $\mathbf{r}$ is an unconstrained convex problem, whose optimal solution can be easily obtained by setting $\frac{\partial f}{\partial\mathbf{r}} = \mathbf{0}$.
The optimal $r_k^\star$ is calculated as
\be\label{eq:update rk}
r_k^\star = \frac{|\mathbf{h}^T_k(\bm{\phi})\mathbf{w}_k|^2}
{\sum_{j\neq k}^{K+M}|\mathbf{h}^T_k(\bm{\phi})\mathbf{w}_j|^2+\sigma_k^2},~\forall k.
\ee
Similarly, the optimal $c_k^\star$ is obtained by setting $\frac{\partial f}{\partial c_k} = 0$ as
\be\label{eq:update ck}
c_k^\star = \frac{\sqrt{1+r_k}\mathbf{h}^T_k(\bm{\phi})\mathbf{w}_k}{\sum_{j=1}^{K+M}|\mathbf{h}^T_k(\bm{\phi})\mathbf{w}_j|^2
+\sigma_k^2},~\forall k.
\ee

\subsubsection{Update $\mathbf{u}$}

Finding $\mathbf{u}$ with the other parameters fixed leads to a feasibility check problem without an explicit objective.
In order to accelerate convergence and leave more DoFs for sum-rate maximization in the next iteration, we propose to update $\mathbf{u}$ by maximizing the SNR lower bound. 
Thus, the optimization problem is formulated as
\be\label{eq:u problem}
\underset{\mathbf{u}}\max~~\frac{L\sigma^2_\text{t}|\mathbf{u}^H(\mathbf{I}_{K+M}\otimes\mathbf{H}_\text{t}(\bm{\phi}))\mathbf{w}|^2}
{\sigma_\text{r}^2\mathbf{u}^H\mathbf{u}},
\ee
which is a typical Rayleigh quotient with the optimal solution
\be\label{eq:update u}
\mathbf{u}^\star = \frac{(\mathbf{I}_{K+M}\otimes\mathbf{H}_\text{t}(\bm{\phi}))\mathbf{w}}
{\mathbf{w}^H(\mathbf{I}_{K+M}\otimes\mathbf{H}^H_\text{t}(\bm{\phi})\mathbf{H}_\text{t}(\bm{\phi}))\mathbf{w}}.
\ee
It is obvious that $\mathbf{u}^\star$ in (\ref{eq:update u}) is feasible to the original feasibility check problem, since the solution obtained in the previous iteration already satisfies the radar SNR constraint, and the solution by maximizing the radar SNR undoubtedly complies with this constraint. 
Moreover, we see that $e^{\jmath\vartheta}\mathbf{u}^\star$ is also an optimal solution to (\ref{eq:u problem}) for an arbitrary angle $\vartheta$, since the phase of the output $\mathbf{u}^H\widetilde{\mathbf{y}}_\text{r}$ does not change the achieved SNR.
Inspired by this finding, after obtaining $\mathbf{u}$ we can restrict the term $\mathbf{u}^H(\mathbf{I}_{K+M}\otimes\mathbf{H}_\text{t}(\bm{\phi}))\mathbf{w}$ to be a non-negative real value, and thus re-formulate the radar SNR constraint (\ref{eq:original problem}b) as
\be
\Re\{\mathbf{u}^H(\mathbf{I}_{K+M}\otimes\mathbf{H}_\text{t}(\bm{\phi}))\mathbf{w}\} \geq \varepsilon_3,
\ee
where for brevity we define $\varepsilon_3 \triangleq \sqrt{\Gamma_\text{t}\sigma_\text{r}^2\mathbf{u}^H\mathbf{u}/(L\sigma^2_\text{t})}$.

\subsubsection{Update $\mathbf{w}$}

With fixed $\mathbf{r}$, $\mathbf{c}$, $\mathbf{u}$, and $\bm{\phi}$, the optimization for the transmit beamforming vector  $\mathbf{w}$ can be formulated as
\begin{subequations}\label{eq:solve W}
\begin{align}
&\underset{\mathbf{w}}\min~~\|\mathbf{B}\mathbf{w}\|^2 -\Re\{\mathbf{a}^H\mathbf{w}\}\\
&~\text{s.t.}~~\Re\{\mathbf{u}^H(\mathbf{I}_{K+M}\otimes\mathbf{H}_\text{t}(\bm{\phi}))\mathbf{w}\} \geq \varepsilon_3,\\
&\quad\quad~\|\mathbf{w}\|^2 \leq P_\text{t}.
\end{align}
\end{subequations}
Obviously, this is a simple convex problem that can be readily solved by various well-developed algorithms or toolboxes \cite{cvx}.

\subsubsection{Update $\bm{\phi}$}

Given $\mathbf{r}$, $\mathbf{c}$, $\mathbf{u}$, and $\mathbf{w}$, the optimization for $\bm{\phi}$ is formulated as
\begin{subequations}\label{eq:solve phi org}
\begin{align}
&\underset{\bm{\phi}}\min~~\bm{\phi}^H\mathbf{D}\bm{\phi}-\Re\{\mathbf{g}^H\bm{\phi}\}\\
&~\text{s.t.}~~\Re\{\mathbf{u}^H(\mathbf{I}_{K+M}\otimes\mathbf{H}_\text{t}(\bm{\phi}))\mathbf{w}\} \geq \varepsilon_3,\label{eq:solve phi orgb}\\
&\quad\quad~|\phi_n| = 1,~~\forall n,\label{eq:solve phi orgc}
\end{align}
\end{subequations}
which cannot be directly solved due to the implicit function with respect to $\bm{\phi}$ in constraint (\ref{eq:solve phi orgb}) and the non-convex unit-modulus constraint (\ref{eq:solve phi orgc}).

We first propose to handle constraint (\ref{eq:solve phi org}b) by re-arranging its left-hand side as an explicit expression with respect to $\bm{\phi}$ and then employing the MM method to find a favorable surrogate function for it.
By recalling the definition of $\mathbf{H}_\text{t}(\bm{\phi})$ and employing the transformations $\bm{\Phi}\mathbf{h}_{\text{r},\text{t}} = \text{diag}\{\mathbf{h}_{\text{r},\text{t}}\}\bm{\phi}$ and $\text{vec}\{\mathbf{ABC}\}=(\mathbf{C}^T\otimes \mathbf{A})\text{vec}\{\mathbf{B}\}$, the term $(\mathbf{I}_{K+M}\otimes\mathbf{H}_\text{t}(\bm{\phi}))\mathbf{w}$ can be equivalently transformed into (\ref{eq:reformulate1b}) presented at the top of the previous page.
Then, constraint (\ref{eq:solve phi org}b) is further re-arranged as
\begin{subequations}\label{eq:new radar constraint}\begin{align}
&\Re\big\{ \mathbf{u}^H(\mathbf{I}\otimes\mathbf{h}_{\text{d},\text{t}}\mathbf{h}_{\text{d},\text{t}}^T)\mathbf{w}
+ \mathbf{u}^H\mathbf{F}\bm{\phi} + \mathbf{u}^H\mathbf{L}\text{vec}\{\bm{\phi}\bm{\phi}^T\}\big\}\\
& = \Re\big\{ \mathbf{u}^H(\mathbf{I}\otimes\mathbf{h}_{\text{d},\text{t}}\mathbf{h}_{\text{d},\text{t}}^T)\mathbf{w}
+ \mathbf{u}^H\mathbf{F}\bm{\phi} + \bm{\phi}^T\widetilde{\mathbf{L}}\bm{\phi}\big\} \geq \varepsilon_3,
\end{align}\end{subequations}
where $\widetilde{\mathbf{L}}\in\mathbb{C}^{N\times N}$ is a reshaped version of $\mathbf{L}^T\mathbf{u}^*$.

Now, it is clear that the third term in (\ref{eq:new radar constraint}b) is a non-concave function, which leads to an intractable constraint.
To solve this problem, we convert the complex-valued function $\Re\{\bm{\phi}^T\widetilde{\mathbf{L}}\bm{\phi}\}$ into the real-valued term $-\overline{\bm{\phi}}^T\overline{\mathbf{L}}\overline{\bm{\phi}}$ by defining $\overline{\bm{\phi}} \triangleq [\Re\{\bm{\phi}^T\}~\Im\{\bm{\phi}^T\}]^T$ and $\overline{\mathbf{L}}\triangleq \bigg[\begin{array}{cc}
-\Re\{\widetilde{\mathbf{L}}\} & \Im\{\widetilde{\mathbf{L}}\} \\
\Im\{\widetilde{\mathbf{L}}\} & \Re\{\widetilde{\mathbf{L}}\}                                                                                    \end{array}\bigg]$, and then employ the MM method to seek a series of tractable surrogate functions for it.
In particular, with the solution $\widehat{\bm{\phi}}$ obtained in the previous iteration, an approximate upper-bound for $\overline{\bm{\phi}}^T\overline{\mathbf{L}}\overline{\bm{\phi}}$ is constructed by using the second-order Taylor expansion as
\begin{subequations}\label{eq:Taylor}\begin{align}
\overline{\bm{\phi}}^T\overline{\mathbf{L}}\overline{\bm{\phi}}
&\leq \widehat{\bm{\phi}}^T\overline{\mathbf{L}}\widehat{\bm{\phi}} \!+\! \widehat{\bm{\phi}}^T(\overline{\mathbf{L}}\!+\!\overline{\mathbf{L}}^T)(\overline{\bm{\phi}}\!-\!\widehat{\bm{\phi}})  \!+\! \frac{\lambda}{2}(\overline{\bm{\phi}}\!-\!\widehat{\bm{\phi}})^T(\overline{\bm{\phi}}\!-\!\widehat{\bm{\phi}})\\
& = \Re\big\{\widehat{\bm{\phi}}^T(\overline{\mathbf{L}}\!+\!\overline{\mathbf{L}}^T\!\!-\!\lambda\mathbf{I}_{2N})\mathbf{U}\bm{\phi}\big\} \!-\!\widehat{\bm{\phi}}^T\overline{\mathbf{L}}^T\!\widehat{\bm{\phi}} \!+\! \lambda N,\!
\end{align}\end{subequations}
where $\lambda$ is the maximum eigenvalue of matrix $(\overline{\mathbf{L}}+\overline{\mathbf{L}}^T)$, $\mathbf{U}\triangleq[\mathbf{I}~~\jmath\mathbf{I}]^H$ converts a real-valued expression into a complex-valued one, and $\overline{\bm{\phi}}^T\overline{\bm{\phi}} = \widehat{\bm{\phi}}^T\widehat{\bm{\phi}} = N$ due to the unit-modulus property of the reflecting coefficients.
Thus, plugging the result in (\ref{eq:Taylor}) into (\ref{eq:new radar constraint}), the radar SNR constraint in each iteration can be concisely re-formulated as
\be\label{eq:radar constraint for phi}
\Re\{\widetilde{\mathbf{u}}^H\bm{\phi}\}\leq \varepsilon_4,
\ee
where we define $\widetilde{\mathbf{u}} \triangleq (-\mathbf{u}^H\mathbf{F} +\widehat{\bm{\phi}}^T(\overline{\mathbf{L}}+\overline{\mathbf{L}}^T-\lambda\mathbf{I}_{2N})\mathbf{U})^H$ and $\varepsilon_4 \triangleq  -\varepsilon_3 + \widehat{\bm{\phi}}^T\overline{\mathbf{L}}^T\widehat{\bm{\phi}} + \Re\{ \mathbf{u}^H(\mathbf{I}\otimes\mathbf{h}_{\text{d},\text{t}}\mathbf{h}_{\text{d},\text{t}}^T)\mathbf{w}\}-\lambda N$.

After converting the radar constraint (\ref{eq:solve phi org}b) to (\ref{eq:radar constraint for phi}), we propose to utilize ADMM to solve for $\bm{\phi}$ under the new constraint (\ref{eq:radar constraint for phi}) and the unit-modulus constraint (\ref{eq:solve phi org}c).
Specifically, an auxiliary variable $\bm{\varphi}\triangleq[\varphi_1,\varphi_2,\ldots,\varphi_N]^T$ is introduced to transform the problem of solving for $\bm{\phi}$ into
\begin{subequations}\label{eq:auxiliary phi}\begin{align}
&\underset{\bm{\phi},\bm{\varphi}}\min~~\bm{\phi}^H\mathbf{D}\bm{\phi}-\Re\{\mathbf{g}^H\bm{\phi}\}\\
&~\text{s.t.}~~\Re\{\widetilde{\mathbf{u}}^H\bm{\phi}\}\leq \varepsilon_4,\\
&\quad\quad~|\phi_n| \leq 1,~~\forall n,\\
&\quad\quad~|\varphi_n| = 1,~~\forall n,\\
&\quad\quad~\bm{\phi} = \bm{\varphi}.
\end{align}
\end{subequations}
Based on ADMM, the solution to (\ref{eq:auxiliary phi}) can be obtained by solving its augmented Lagrangian function:
\begin{subequations}\label{eq:solve for phi and variphi}
\begin{align}
&\underset{\bm{\phi},\bm{\varphi}}\min~~\bm{\phi}^H\mathbf{D}\bm{\phi}-\Re\{\mathbf{g}^H\bm{\phi}\} + \frac{1}{2\rho }\|\bm{\phi}-\bm{\varphi}+\rho\bm{\mu}\|^2\\
&~\text{s.t.}\quad~(\ref{eq:auxiliary phi}\text{b})-(\ref{eq:auxiliary phi}\text{d}),
\end{align}
\end{subequations}
where $\bm{\mu}\in\mathbb{C}^N$ is the dual variable and $\rho>0$ is a pre-set penalty parameter.
This multivariate problem can be solved by alternately updating each variable given the others.

\textbf{Update $\bm{\phi}$}:
It is obvious that with fixed $\bm{\varphi}$ and $\bm{\mu}$, the optimization problem for updating $\bm{\phi}$ is convex and can be readily solved by various existing efficient algorithms.

\textbf{Update $\bm{\varphi}$}:
Given $\bm{\phi}$ and $\bm{\mu}$, the optimal $\bm{\varphi}^\star$ can be easily obtained by phase alignment
\be\label{eq:update varphi}
\bm{\varphi}^\star = e^{\jmath\angle(\bm{\phi}+\rho\bm{\mu})}.
\ee

\textbf{Update $\bm{\mu}$}:
After obtaining $\bm{\phi}$ and $\bm{\varphi}$, the dual variable $\bm{\mu}$ is updated by
\be\label{eq:update mu}
\bm{\mu}:=\bm{\mu} +(\bm{\phi}-\bm{\varphi})/\rho.
\ee

\begin{algorithm}[!t]
	\begin{small}
		\caption{SNR-Constrained Joint Beamforming and Reflection Design}
		\label{alg}
		\begin{algorithmic}[1]
			\REQUIRE $\mathbf{h}_{\text{d},\text{t}}$, $\mathbf{h}_{\text{r},\text{t}}$, $\mathbf{G}$, $\sigma_\text{t}^2$, $\sigma_\text{r}^2$, $\mathbf{h}_{\text{d},k}$, $\mathbf{h}_{\text{r},k}$, $\sigma_k^2$, $\forall k$, $P_\text{t}$, $L$, $\Gamma_\text{t}$, $\rho$.
			\ENSURE $\mathbf{W}^\star$, $\bm{\phi}^\star$, and $\mathbf{u}^\star$.
			\STATE {Initialize $\bm{\phi}$ and $\mathbf{W}$ using RCG.}
			\WHILE {no convergence }
			\STATE{Update $r_k,~\forall k$, by (\ref{eq:update rk}).}
			\STATE{Update $c_k,~\forall k$, by (\ref{eq:update ck}).}
			\STATE{Update $\mathbf{u}$ by (\ref{eq:update u}).}
			\STATE{Update $\mathbf{w}$ by solving problem (\ref{eq:solve W}).}
			\STATE{Update $\bm{\phi}$ by solving problem (\ref{eq:solve for phi and variphi}) given other variables.}
			\STATE{Update $\bm{\varphi}$ by (\ref{eq:update varphi}).}
			\STATE{Update $\bm{\mu}$ by (\ref{eq:update mu}).}
			\STATE{$\rho:=0.8\rho$.}
			\ENDWHILE
			\STATE{Reshape $\mathbf{w}$ to $\mathbf{W}$.}
			\STATE{Return $\mathbf{W}^\star = \mathbf{W}$, $\bm{\phi}^\star = \bm{\phi}$, and $\mathbf{u}^\star = \mathbf{u}$.}
		\end{algorithmic}
	\end{small}
\end{algorithm}
\subsection{Summary and Initialization}\label{sec:summary and initialize}

Based on the above derivations, the proposed SNR-constrained joint beamforming and reflection design is straightforward and summarized in Algorithm 1.
With an appropriate initialization, we iteratively update each variable until convergence.
It is noted that the penalty parameter $\rho$ is shrunk in each iteration to force the equality constraint to be satisfied. 
As the penalty parameter gradually decreases, i.e., $\rho\rightarrow 0$, the solution to the problem (\ref{eq:solve for phi and variphi}) ultimately ensures that the unit-modulus constraint is satisfied. 
While for any fixed $\rho$, we note that the achieved objective value of problem (\ref{eq:solve for phi and variphi}) is an upper bound of that achieved by the optimal solution to the problem (\ref{eq:solve phi org}). 
By alternately solving the problem (\ref{eq:solve for phi and variphi}), the upper bound can be gradually tightened.
Since each sub-problem of the problem (\ref{eq:solve for phi and variphi}) is optimally solved, the objective function (\ref{eq:solve for phi and variphi}a) is monotonically non-increasing, and the solution acquired via alternative optimization can ensure convergence to a stationary point of the problem (\ref{eq:solve phi org}).
In addition, we see that the original problem (\ref{eq:original problem}) is solved by alternately updating $\mathbf{r}$, $\mathbf{c}$, $\mathbf{u}$, $\mathbf{w}$, and $\bm{\phi}$, whose stationary points are achieved respectively. 
Note that the objective value (\ref{eq:original problem}a) is non-decreasing over iterations, and any limit point of $\{ \mathbf{r}, \mathbf{c}, \mathbf{u}, \mathbf{w}, \bm{\phi}\}$ is a stationary point of the original optimization problem (\ref{eq:original problem}).
Furthermore, the objective value of (\ref{eq:original problem}a) is upper bounded by a finite value owing to the transmit power budget. 
Therefore, Algorithm 1 guarantees convergence to a stationary point and a locally optimal solution.

Since the starting point is important for the proposed alternating algorithm, we investigate some straightforward methods for appropriately initializing $\bm{\phi}$ and $\mathbf{w}$.
Intuitively, the RIS is deployed for the purpose of improving the quality of the propagation environment between the BS and the target/communication users, and the channel gain can be regarded as a metric for channel quality.
Therefore, we propose to initialize $\bm{\phi}$ by maximizing the channel gains of the target and the communication users, which can be formulated as a typical problem in the literature of RIS and then efficiently solved using the Riemannian conjugate gradient (RCG) algorithm \cite{Liu TWC 2021}.
After initializing $\bm{\phi}$, we propose to initialize $\mathbf{w}$ by using the available transmit power to maximize the sum power of the received signals of the target and the communication users, which can also be solved using the RCG algorithm.

Finally, we briefly analyze the computational complexity of the proposed SNR-constrained joint beamforming design algorithm. As shown in Algorithm 1, the computational burden mainly results from the update for $\mathbf{w}$ and $\bm{\phi}$. We assume that the popular interior point method is utilized to solve these convex sub-problems. Thus, the complexities of updating $\mathbf{w}$ and $\bm{\phi}$ is of order $\mathcal{O}\{M^{3.5}(M+K)^{3.5}\}$ and $\mathcal{O}\{N^{3.5}\}$, respectively, and the overall complexity is of order $\mathcal{O}\{M^{3.5}(M+K)^{3.5}+N^{3.5}\}$.

\section{CRB-Constrained Joint Beamforming and Reflection Design}\label{sec: CRB constrained design}

In this section, we focus on the CRB-constrained joint beamforming and reflection design to ensure the parameter estimation performance.
In particular, we investigate optimizing $\mathbf{W}$ and $\bm{\phi}$ to maximize the sum-rate while satisfying a CRB constraint, the transmit power budget and the unit-modulus constraint.
The optimization problem is thus formulated as
\begin{subequations}\label{eq:CRB constrained prob}\begin{align}
&\underset{\mathbf{W},\bm{\phi}}\max~~\sum_{k=1}^K\log_2(1+\text{SINR}_k)\\
&~~\text{s.t.}\quad\text{Tr}\big\{(\mathbf{F}_{\bm{\theta}\bm{\theta}^T}-\mathbf{F}_{\bm{\theta}\bm{\alpha}^T}
\mathbf{F}_{\bm{\alpha}\bm{\alpha}^T}^{-1}\mathbf{F}^T_{\bm{\theta}\bm{\alpha}^T})^{-1}\big\} \leq \varepsilon, \\
&\quad\quad\quad\|\mathbf{W}\|_F^2 \leq P_\text{t},\\
&\quad\quad\quad|\phi_n| = 1,~~\forall n,
\end{align}
\end{subequations}
where $\varepsilon$ is the CRB threshold.
The same FP-based procedure in Sec. IV-A can be utilized to transform the objective function into $\mathcal{F}(\mathbf{w},\bm{\phi},\mathbf{r},\mathbf{c})$ as in (\ref{eq: new obj func}), which is a concave function with respect to each variable.
Thus, the CRB constraint (\ref{eq:CRB constrained prob}b) is the major difficulty that requires sophisticated derivations and transformations to facilitate the algorithm development.

\subsection{CRB Constraint Transformation}

To handle the CRB constraint, we first introduce an auxiliary variable $\mathbf{J}\in\mathbb{C}^{2\times2}$, $\mathbf{J}\succeq \mathbf{0}$.
Since the matrix $\mathbf{F}_{\bm{\theta}\bm{\theta}^T}-\mathbf{F}_{\bm{\theta}\bm{\alpha}^T}
\mathbf{F}_{\bm{\alpha}\bm{\alpha}^T}^{-1}\mathbf{F}^T_{\bm{\theta}\bm{\alpha}^T}$ is positive semidefinite and the function $\text{Tr}\{\mathbf{A}^{-1}\}$ is decreasing on the space of positive semidefinite matrices, the CRB constraint (\ref{eq:CRB constrained prob}b) can be converted into the following two constraints \cite{Wang-arXiv-2022}:
\begin{subequations}\label{eq:new CRB constraint}\begin{align}
&\text{Tr}\{\mathbf{J}^{-1}\} \leq \varepsilon, \\
&\mathbf{F}_{\bm{\theta}\bm{\theta}^T}-\mathbf{F}_{\bm{\theta}\bm{\alpha}^T}
\mathbf{F}_{\bm{\alpha}\bm{\alpha}^T}^{-1}\mathbf{F}^T_{\bm{\theta}\bm{\alpha}^T} \succeq \mathbf{J}.
\end{align}\end{subequations}
Then, by applying the Schur complement \cite{Shur complement}, we can re-formulate (\ref{eq:new CRB constraint}b) as
\be\label{eq:schur CRB}
\left[\begin{array}{cc}\mathbf{F}_{\bm{\theta}\bm{\theta}^T}-\mathbf{J} & \mathbf{F}_{\bm{\theta}\bm{\alpha}^T}\\ \mathbf{F}_{\bm{\theta}\bm{\alpha}^T}^T & \mathbf{F}_{\bm{\alpha}\bm{\alpha}^T}\end{array}\right] \succeq \mathbf{0}.
\ee
We see that constraint (\ref{eq:new CRB constraint}a) is convex, while constraint (\ref{eq:schur CRB}) is still very difficult to tackle since the variables $\mathbf{W}$ and $\bm{\phi}$ are non-linearly coupled in each element of the positive semidefinite matrix.
To solve this problem, we introduce an auxiliary variable $\mathbf{f}\triangleq [f_1,f_2,f_3,f_4,f_5,f_6]^T$ to take $\mathbf{W}$ and $\bm{\phi}$ out of the positive semidefinite matrix constraint and transform the problem to
\begin{subequations}\label{eq:opt1}\begin{align}
&\underset{\mathbf{w},\bm{\phi},\mathbf{r},\mathbf{c},\mathbf{J},\mathbf{f}}\max~~\mathcal{F}(\mathbf{w},\bm{\phi},\mathbf{r},\mathbf{c})\\
&\quad~\text{s.t.}\qquad\text{Tr}\{\mathbf{J}^{-1}\} \leq \varepsilon, ~~\mathbf{J}\succeq \mathbf{0},\\
&~~\left[\!\!\begin{array}{cc}\frac{2L|\alpha_\text{t}|^2}{\sigma_\text{r}^2}\Re\!\left\{\!\left[\!\!\begin{array}{cc}
f_1 & f_2\\f_2 &f_4\end{array}\!\!\right]\!\right\}\!-\!\mathbf{J} & \frac{2L}{\sigma_\text{r}^2}\Re\!\left\{\!\alpha_\text{t}^*\!\left[\!\!\begin{array}{c}
f_3\\f_5\end{array}\!\!\right][1~\jmath]\right\} \\ \frac{2L}{\sigma_\text{r}^2}\Re\left\{\alpha_\text{t}^*\left[\begin{array}{c}
1\\ \jmath\end{array}\right][f_3~f_5]\right\} & \frac{2L}{\sigma_\text{r}^2}f_6\mathbf{I}_2\end{array}\!\!\!\right] \succeq \mathbf{0}, \\
&\quad\|\mathbf{w}\|_2^2 \leq P_\text{t},\\
&\quad|\phi_n| = 1,~~\forall n,\\
&\quad f_i = \mathcal{F}_i(\mathbf{W}, \bm{\phi}),~\forall i,
\end{align}
\end{subequations}
where $\mathcal{F}_i(\mathbf{W}, \bm{\phi}),~\forall i$ are functions with respect to $\mathbf{W}$ and $\bm{\phi}$ according to the definition of $\mathbf{F}_\text{IM}$ in (\ref{eq:FIM partition}) and (\ref{eq:each ele of FIM}).
Detailed expressions for them are presented in Appendix B, where we re-arrange them into explicit forms with respect to $\mathbf{W}$ and $\bm{\phi}$ for the following algorithm development.

Then, the augmented Lagrangian function of problem (\ref{eq:opt1}) is formulated as
\begin{subequations}\label{eq:AL prob for CRB}\begin{align}
&\underset{\mathbf{w},\bm{\phi},\mathbf{r},\mathbf{c},\mathbf{J},\mathbf{f}}\min-\mathcal{F}(\mathbf{w},\bm{\phi},\mathbf{r},\mathbf{c})
+\frac{1}{2\rho_1}\sum_{i=1}^6|\mathcal{F}_i(\mathbf{W}, \bm{\phi})\!-\!f_i\!+\!\rho_1\zeta_i|^2\\
&\quad~\text{s.t.}\qquad(\ref{eq:opt1}\text{b})-(\ref{eq:opt1}\text{e}),
\end{align}
\end{subequations}
where $\bm{\zeta}\triangleq[\zeta_1,\zeta_2,\ldots,\zeta_6]^T$ is the dual variable and $\rho_1>0$ is the penalty parameter.
Now we can use the block coordinate descent (BCD) method to alternately update each variable as presented in what follows.

\subsection{Block Update}

\subsubsection{Update $\mathbf{r}$ and $\mathbf{c}$}

Fixing other variables, the optimal solution for the auxiliary variables $\mathbf{r}$ and $\mathbf{c}$ introduced based on FP can be obtained as in (\ref{eq:update rk}) and (\ref{eq:update ck}), respectively.

\subsubsection{Update $\mathbf{J}$ and $\mathbf{f}$}

Given other variables, the optimization problem for updating $\mathbf{J}$ and $\mathbf{f}$ can be formulated as
\begin{subequations}\label{eq:solve for J f}\begin{align}
&\underset{\mathbf{J},\mathbf{f}}\min~~\sum_{i=1}^6|\mathcal{F}_i(\mathbf{W}, \bm{\phi})-f_i+\rho_1\zeta_i|^2\\
&~\text{s.t.}\quad(\ref{eq:opt1}\text{b}),~(\ref{eq:opt1}\text{c}).
\end{align}\end{subequations}
We observe that this is a semidefinite programming (SDP) problem, which can be readily solved by standard algorithms.

\subsubsection{Update $\mathbf{w}$}

Substituting the definitions of $\mathcal{F}(\mathbf{w},\bm{\phi},\mathbf{r},\mathbf{c})$ in (\ref{eq: new obj func}) and $\mathcal{F}_i(\mathbf{W}, \bm{\phi})$ in (\ref{eq:ftilde}) into (\ref{eq:AL prob for CRB}), the sub-problem of solving for $\mathbf{w}$ can be expressed as
\begin{subequations}\label{eq:sub prob of w for CRB}\begin{align}
&\underset{\mathbf{w}}\min~\|\mathbf{Bw}\|^2-\Re\{\mathbf{a}^H\mathbf{w}\}+\frac{1}{2\rho_1}\sum_{i=1}^6
|\text{Tr}\{\mathbf{A}_i\mathbf{WW}^H\}\!+\!d_i|^2\\
&~\text{s.t.}~~\|\mathbf{w}\|_2^2 \leq P_\text{t},
\end{align}\end{subequations}
where we define the constant term $d_i \triangleq -f_i+\rho_1\zeta_i$ for notational simplicity and $\mathbf{A}_i,~\forall i$ in Appendix B.
It is obvious that the third term in the objective function (\ref{eq:sub prob of w for CRB}a) is non-convex and quartic with respect to $\mathbf{W}$, which greatly hinders the solution.
To tackle this difficulty, we utilize the MM method to find a series of tractable surrogate functions. 

Specifically, we first expand the third non-convex quartic term in (\ref{eq:sub prob of w for CRB}a) as
\begin{subequations}\label{eq:trAWWh}\begin{align}
&|\text{Tr}\{\mathbf{A}_i\mathbf{WW}^H\}\!+\!d_i\}|^2 = |\mathbf{w}^H(\mathbf{I}_{K+M}\!\otimes\!\mathbf{A}_i)\mathbf{w} + d_i|^2 \\
& = |d_i|^2 + 2\Re\{d_i^*\mathbf{w}^H(\mathbf{I}\!\otimes\!\mathbf{A}_i)\mathbf{w}\}
+ |\mathbf{w}^H(\mathbf{I}\!\otimes\!\mathbf{A}_i)\mathbf{w}|^2,
\end{align}\end{subequations}
in which we utilize the transformation $\text{Tr}\{\mathbf{ABCD}\} = \text{vec}^H\{\mathbf{D}^H\}(\mathbf{C}^T\otimes\mathbf{A})\text{vec}\{\mathbf{B}\}$ to obtain (\ref{eq:trAWWh}a).
Then, we derive tractable surrogation functions for the second and third terms of (\ref{eq:trAWWh}b).
A convex surrogate function for the second term of (\ref{eq:trAWWh}b) is derived as
\begin{subequations}\label{eq:second term}\begin{align}
&2\Re\{d_i^*\mathbf{w}^H(\mathbf{I}\!\otimes\!\mathbf{A}_i)\mathbf{w}\} =\mathbf{w}^H\widetilde{\mathbf{A}}_i\mathbf{w}\\
&\qquad\qquad\qquad\quad\leq 2\Re\{\mathbf{w}^H(\widetilde{\mathbf{A}}_i\!-\!\lambda_{\text{t},i}\mathbf{I})\mathbf{w}^t\} + \beta_{1,i},
\end{align}\end{subequations}
where we define $\widetilde{\mathbf{A}}_i \triangleq d_i^*\mathbf{I}\otimes\mathbf{A}_i + d_i\mathbf{I}\otimes\mathbf{A}_i^H$ to drop the real operation.
In (\ref{eq:second term}b), a second-order Taylor expansion and the power constraint $\|\mathbf{w}\|_2^2 \leq P_\text{t}$ are applied, and $\lambda_{\text{t},i}$ denotes the maximum eigenvalue of $\widetilde{\mathbf{A}}_i$. According to the properties of Kronecker product, $\lambda_{\text{t},i}$ equals to the maximum eigenvalue of the matrix $(d_i^*\mathbf{A}_i + d_i\mathbf{A}_i^H)$.
In addition, $\mathbf{w}^t$ represents the obtained solution in the $t$-th iteration, and the constant term $\beta_{1,i}$ is defined as $\beta_{1,i} \triangleq \lambda_{\text{t},i}P_\text{t} + (\mathbf{w}^t)^H(\lambda_{\text{t},i}\mathbf{I} - \widetilde{\mathbf{A}}_i)\mathbf{w}^t$.

Similarly, a convex surrogate function for the third term of (\ref{eq:trAWWh}b) can be derived as
\begin{subequations}\label{eq:w4}\begin{align}
& |\mathbf{w}^H(\mathbf{I}\otimes\mathbf{A}_i)\mathbf{w}|^2 \\
&= \text{Tr}\{(\mathbf{I}\otimes\mathbf{A}_i)\mathbf{w}
\mathbf{w}^H(\mathbf{I}\otimes\mathbf{A}_i^H)\mathbf{w}\mathbf{w}^H\} = \overline{\mathbf{w}}^H\overline{\mathbf{A}}_i\overline{\mathbf{w}}\\
& \leq \Re\{(\overline{\mathbf{w}}^t)^H(\overline{\mathbf{A}}_i+\overline{\mathbf{A}}_i^H-2\lambda_{\text{b},i}\mathbf{I})\overline{\mathbf{w}}\}
+ \beta_{2,i} \\
& = |\mathbf{w}^H\text{vec}\{\mathbf{A}_i\mathbf{W}^t\}|^2 + |\mathbf{w}^H\text{vec}\{\mathbf{A}_i^H\mathbf{W}^t\}|^2 \non\\
&\quad - 2\lambda_{\text{b},i}\mathbf{w}^H\mathbf{w}^t(\mathbf{w}^t)^H\mathbf{w} + \beta_{2,i},
\end{align}\end{subequations}
where we define $\overline{\mathbf{w}}\triangleq \text{vec}\{\mathbf{ww}^H\}$ and $\overline{\mathbf{A}}_i\triangleq(\mathbf{I}\otimes\mathbf{A}_i^*)\otimes(\mathbf{I}\otimes\mathbf{A}_i)$.
In (\ref{eq:w4}c), we apply a second-order Taylor expansion and the power constraint (\ref{eq:opt1}d), denote $\lambda_{\text{b},i}$ as the maximum eigenvalue of $\overline{\mathbf{A}}_i$, which equals $\max\{|\lambda_{i,j}|^2,~\forall j\}$ where $\lambda_{i,j}$ represents the $j$-th eigenvalue of $\mathbf{A}_i$, and define $\beta_{2,i} \triangleq \lambda_{\text{b},i}P_\text{t}^2 + \lambda_{\text{b},i}(\overline{\mathbf{w}}^t)^H\overline{\mathbf{w}}^t
- (\overline{\mathbf{w}}^t)^T\overline{\mathbf{A}}_i^T(\overline{\mathbf{w}}^t)^*$.
We see that only the third term of (\ref{eq:w4}d) is non-convex, whose convex surrogate function is derived using a first-order Taylor expansion as
\be\label{eq:use first order}
\mathbf{w}^H\mathbf{w}^t(\mathbf{w}^t)^H\mathbf{w} \geq \|\mathbf{w}^t\|^4 \!+\! 2\Re\{\|\mathbf{w}^t\|^2(\mathbf{w}^t)^H(\mathbf{w}\!-\!\mathbf{w}^t)\}.\ee

Substituting the results obtained in (\ref{eq:trAWWh})-(\ref{eq:use first order}) into (\ref{eq:sub prob of w for CRB}a), the sub-problem for updating $\mathbf{w}$ can be written as
\begin{subequations}\label{eq:solve for w}\begin{align}
&\underset{\mathbf{w}}\min~\|\mathbf{Bw}\|^2-\Re\{\mathbf{a}^H\mathbf{w}\}+\frac{1}{2\rho_1}
\sum_{i=1}^6(|\mathbf{w}^H\text{vec}\{\mathbf{A}_i\mathbf{W}^t\}|^2 \non \\
&\qquad + |\mathbf{w}^H\text{vec}\{\mathbf{A}_i^H\mathbf{W}^t\}|^2 + \Re\{\widetilde{\mathbf{a}}_i^H\mathbf{w}\} )\\
&~\text{s.t.}~~\|\mathbf{w}\|_2^2 \leq P_\text{t},
\end{align}\end{subequations}
where for simplicity we define $\widetilde{\mathbf{a}}_i^H \triangleq2(\mathbf{w}^t)^H
(\widetilde{\mathbf{A}}_i-\lambda_{\text{t},i}\mathbf{I}-2\lambda_{\text{b},i}\|\mathbf{w}^t\|^2\mathbf{I})$.
It is clear that problem (\ref{eq:solve for w}) is convex and can be readily solved by standard convex optimization algorithms.

\subsubsection{Update $\bm{\phi}$}

Fixing other variables, the sub-problem of solving for $\bm{\phi}$ is formulated as
\begin{subequations}\label{eq:subproblem with fixed W}\begin{align}
&\underset{\bm{\phi}}\min~\bm{\phi}^H\mathbf{D}\bm{\phi}-\Re\{\mathbf{g}^H\bm{\phi}\}
+\frac{1}{2\rho_1}\sum_{i=1}^6|\mathcal{F}_i(\mathbf{W},\bm{\phi})+d_i|^2\\
&~\text{s.t.}~~|\phi_n| = 1,~~\forall n.
\end{align}
\end{subequations}
As given in (\ref{eq:ftilde}), $\mathcal{F}_i(\mathbf{W},\bm{\phi}),~\forall i$ are complicated and non-convex functions with respect to $\bm{\phi}$, which makes the third term in (\ref{eq:subproblem with fixed W}a) very difficult to handle.
To solve this problem, we introduce three auxiliary variables $\bm{\varphi}\triangleq[\varphi_1,\varphi_2,\ldots,\varphi_N]^T$, $\mathbf{v}\triangleq[v_1,v_2,\ldots,v_{N^2}]^T$ and $\bm{\nu}\triangleq[\nu_1,\nu_2,\ldots,\nu_{N^2}]^T$, and convert $\mathcal{F}_i(\mathbf{W},\bm{\phi})$ into linear functions with respect to each variable:
\begin{subequations}\label{eq:linear ftilde}\begin{align}
\mathcal{F}_1(\bm{\phi},\bm{\varphi},\mathbf{v},\bm{\nu}) &= q_1 + \bm{\phi}^H\mathbf{u}_1 + \mathbf{u}_1^H\bm{\varphi} + \bm{\phi}^H\mathbf{U}_1\bm{\varphi},\\
\mathcal{F}_2(\bm{\phi},\bm{\varphi},\mathbf{v},\bm{\nu}) &= \overline{\mathbf{u}}_2^H\bm{\varphi} + \bm{\phi}^H\mathbf{U}_2\bm{\varphi} \!+\! \overline{\mathbf{z}}_2^H\bm{\nu} \!+\! \bm{\phi}^H\mathbf{C}_2\bm{\nu},\\
\mathcal{F}_3(\bm{\phi},\bm{\varphi},\mathbf{v},\bm{\nu}) &= q_3+\bm{\phi}^H\mathbf{u}_3+ \overline{\mathbf{u}}_3^H\bm{\varphi}+ \bm{\phi}^H\mathbf{U}_3\bm{\varphi}  \non\\
&\quad + \overline{\mathbf{z}}_3^H\bm{\nu} + \bm{\phi}^H\mathbf{C}_3\bm{\nu},\\
\mathcal{F}_4(\bm{\phi},\bm{\varphi},\mathbf{v},\bm{\nu}) &= \bm{\phi}^H\mathbf{U}_4\bm{\varphi} + \mathbf{v}^H\overline{\mathbf{C}}_4\bm{\varphi} \!+\! \bm{\phi}^H\overline{\mathbf{C}}_4^H\bm{\nu} \!+\! \mathbf{v}^H\mathbf{Z}_4\bm{\nu},\\
\mathcal{F}_5(\bm{\phi},\bm{\varphi},\mathbf{v},\bm{\nu}) &= \bm{\phi}^H\mathbf{u}_5 + \bm{\phi}^H\mathbf{U}_5\bm{\varphi} + \mathbf{v}^H\mathbf{z}_5 +
\mathbf{v}^H\mathbf{Z}_5\bm{\nu}\non\\
 &\quad + \mathbf{v}^H\overline{\mathbf{C}}_5\bm{\varphi} + \bm{\phi}^H\mathbf{C}_5\bm{\nu},\\
\mathcal{F}_6(\bm{\phi},\bm{\varphi},\mathbf{v},\bm{\nu}) &=q_6 + \bm{\phi}^H\mathbf{u}_6 + \mathbf{u}_6^H\bm{\varphi} \!+\! \bm{\phi}^H\mathbf{U}_6\bm{\varphi} \!+\! \mathbf{v}^H\overline{\mathbf{C}}_6\bm{\varphi} \non\\
&\quad  +\!   \mathbf{z}_6^H\bm{\nu} \!+\! \mathbf{v}^H\mathbf{z}_6 \!+\!  \bm{\phi}^H\overline{\mathbf{C}}_6^H\bm{\nu}\!+\!  \mathbf{v}^H\mathbf{Z}_6\bm{\nu}.\!
\end{align}\end{subequations}
Then, the optimization problem for finding these variables can be formulated as
\begin{subequations}\label{eq:solve for phi etc}\begin{align}
&\underset{\bm{\phi},\bm{\varphi},\mathbf{v},\bm{\nu}}\min~\bm{\phi}^H\mathbf{D}\bm{\phi}\!-\!\Re\{\mathbf{g}^H\bm{\phi}\}
\!+\!\frac{1}{2\rho_1}\sum_{i=1}^6|\mathcal{F}_i(\bm{\phi},\bm{\varphi},\mathbf{v},\bm{\nu})\!+\!d_i|^2\\
&~\text{s.t.}~~\bm{\varphi} = \bm{\phi},\quad\mathbf{v} = \bm{\phi}\otimes\bm{\varphi},\quad\bm{\nu} = \mathbf{v},\\
&\qquad~|\varphi_n| = |\phi_n| = 1,~~\forall n,\quad |\nu_j| =|v_j| = 1,~~\forall j.
\end{align}
\end{subequations}
By treating equality constraints (\ref{eq:solve for phi etc}b) as penalty terms, the augmented Lagrangian function of problem (\ref{eq:solve for phi etc}) is written as
\begin{subequations}\label{eq:AL function for phi}\begin{align}
&\underset{\bm{\phi},\bm{\varphi},\mathbf{v},\bm{\nu}}\min~\bm{\phi}^H\mathbf{D}\bm{\phi}\!-\!\Re\{\mathbf{g}^H\bm{\phi}\}
\!+\!\frac{1}{2\rho_1}\sum_{i=1}^6|\mathcal{F}_i(\bm{\phi},\bm{\varphi},\mathbf{v},\bm{\nu})\!+\!d_i|^2 \non\\
&\hspace{1 cm} +
\frac{1}{2\rho_2}\|\bm{\phi}-\bm{\varphi}+\rho_2\bm{\mu}\|^2 + \frac{1}{2\rho_3}\|\mathbf{v}-\bm{\phi}\otimes\bm{\varphi}+\rho_3\bm{\lambda}\|^2 \non\\
&\hspace{1 cm}+\frac{1}{2\rho_4}\|\bm{\nu}-\mathbf{v}+\rho_4\bm{\omega}\|^2 \\
&~~\text{s.t.}\quad|\varphi_n| = |\phi_n| = 1,~~\forall n,\quad|\nu_j| =|v_j| = 1,~~\forall j,
\end{align}
\end{subequations}
where $\bm{\mu}\in\mathbb{C}^N$, $\bm{\lambda}\in\mathbb{C}^{N^2}$ and $\bm{\omega}\in\mathbb{C}^{N^2}$ are dual variables, and $\rho_2$, $\rho_3$ and $\rho_4$ are penalty parameters.
Thanks to the definitions in (\ref{eq:linear ftilde}), the term $\mathcal{F}_i(\bm{\phi},\bm{\varphi},\mathbf{v},\bm{\nu}) +d_i$ can be re-arranged as linear functions with respect to each variable as
\begin{subequations}\label{eq:ftilde wrt each}\begin{align}
\mathcal{F}_i(\bm{\phi},\bm{\varphi},\mathbf{v},\bm{\nu}) +d_i &= a_{\phi,i}+\bm{\phi}^H\mathbf{b}_{\phi,i} = a_{\varphi,i}+\mathbf{b}_{\varphi,i}^H\bm{\varphi}\\
&= a_{v,i}+\mathbf{v}^H\mathbf{b}_{v,i} = a_{\nu,i}+\mathbf{b}_{\nu,i}^H\bm{\nu},
\end{align}\end{subequations}
where the definitions of $a_{\phi,i}$, $\mathbf{b}_{\phi,i}$, $a_{\varphi,i}$, $\mathbf{b}_{\varphi,i}$, $a_{v,i}$, $\mathbf{b}_{v,i}$, $a_{\nu,i}$, and $\mathbf{b}_{\nu,i}$ are straightforward based on (\ref{eq:linear ftilde}).
In addition, the term $\bm{\phi}\otimes\bm{\varphi}$ can be re-arranged as
\begin{equation}\label{eq:kron term}\begin{aligned}
	\bm{\phi}\otimes\bm{\varphi} &= \text{vec}\{\bm{\varphi}\bm{\phi}^T\} \\
	&= (\bm{\phi}\otimes\mathbf{I}_N)\bm{\varphi} \\
	&= (\mathbf{I}_N\otimes\bm{\varphi})\bm{\phi}.
\end{aligned}\end{equation}
Plugging the results of (\ref{eq:ftilde wrt each}) and (\ref{eq:kron term}) into (\ref{eq:AL function for phi}), we can explicitly formulate the sub-problem for each variable as follows.

\textbf{Update $\bm{\phi}$}:
The sub-problem of updating $\bm{\phi}$ is written as
\begin{subequations}\label{eq:subsub prob for phi}\begin{align}
&\underset{\bm{\phi}}\min~~\bm{\phi}^H{\mathbf{D}}_\phi\bm{\phi}+\Re\{{\mathbf{g}}_\phi^H\bm{\phi}\}  \\
&~\text{s.t.}~~|\phi_n|=1,~\forall n,
\end{align}
\end{subequations}
where we define
\begin{subequations}\label{eq:define Dphi}\begin{align}
{\mathbf{D}}_\phi &\triangleq \mathbf{D}+\frac{1}{2\rho_1}\sum_{i=1}^6\mathbf{b}_{\phi,i}\mathbf{b}_{\phi,i}^H,\\ {\mathbf{g}}_\phi&\triangleq \frac{1}{\rho_1}\sum_{i=1}^6a_{\phi,i}^*\mathbf{b}_{\phi,i} + \bm{\mu} - \frac{\bm{\varphi}}{\rho_2}
-(\mathbf{I}_N\!\otimes\!\bm{\varphi}^H)(\bm{\lambda}+\frac{\mathbf{v}}{\rho_3})-\mathbf{g}.
\end{align}
\end{subequations}
This is a typical optimization problem in the literature of RIS. 
Considering that the popular semi-definite relaxation (SDR) \cite{Wu-TWC-2019} may suffer from high complexity and the alternating-based algorithms such as RCG and element-wise BCD require additional iterative loops, a direct closed-form solution is preferred especially for large $N$. 
Therefore, we rely on the MM method to seek a favorable surrogate objective function whose optimal solution under the unit-modulus constraint can be easily obtained in closed form.

Particularly, we observe that matrix $\mathbf{D}_\phi$ is positive-definite and it is the summation of several rank-one matrices.
In addition, using the second-order Taylor expansion, a linear surrogate function for the term $\bm{\phi}^H\mathbf{b}\mathbf{b}^H\bm{\phi}$ can be derived as
\be\begin{aligned}\label{eq:MM for phi}
\bm{\phi}^H\mathbf{b}\mathbf{b}^H\bm{\phi} &\leq N\|\mathbf{b}\|^2 +
2\Re\{(\bm{\phi}^t)^H(\mathbf{b}\mathbf{b}^H\!-\!\|\mathbf{b}\|^2\mathbf{I}_N)\bm{\phi}\} \\
&\quad + (\bm{\phi}^t)^H(\|\mathbf{b}\|^2\mathbf{I}_N-\mathbf{b}\mathbf{b}^H)\bm{\phi}^t,
\end{aligned}\ee
where $\bm{\phi}^t$ represents the solution obtained in the $t$-th iteration and we use the fact that the maximum eigenvalue of $\mathbf{bb}^H$ equals $\|\mathbf{b}\|^2$ and $(\bm{\phi}^t)^H\bm{\phi}^t=N$ due to the unit-modulus constraint (\ref{eq:subsub prob for phi}b).
Based on the result in (\ref{eq:MM for phi}), a linear surrogate function for $\bm{\phi}^H{\mathbf{D}}_\phi\bm{\phi}$ can be obtained as
\be\label{eq:phi MM}\begin{aligned}\bm{\phi}^H{\mathbf{D}}_\phi\bm{\phi} &\leq N\|{\mathbf{D}}_\phi\|_F^2 + 2\Re\{(\bm{\phi}^t)^H({\mathbf{D}}_\phi-\|{\mathbf{D}}_\phi\|_F^2\mathbf{I}_N)\bm{\phi}\} \\
&\quad +
(\bm{\phi}^t)^H(\|{\mathbf{D}}_\phi\|_F^2\mathbf{I}_N-{\mathbf{D}}_\phi)\bm{\phi}^t.
\end{aligned}\ee
Substituting (\ref{eq:phi MM}) into (\ref{eq:subsub prob for phi}a), the problem of solving for $\bm{\phi}$ can be written as
\begin{subequations}\label{eq:surrogate prob for phi}\begin{align}
&\underset{\bm{\phi}}\min~~\Re\{\widetilde{\mathbf{g}}_\phi^H\bm{\phi}\}  \\
&~\text{s.t.}~~|\phi_n|=1,~\forall n,
\end{align}
\end{subequations}
where we define $\widetilde{\mathbf{g}}_\phi \triangleq \mathbf{g}_\phi + 2({\mathbf{D}}_\phi-\|{\mathbf{D}}_\phi\|_F^2\mathbf{I})\bm{\phi}^t$ for simplicity.
The optimal solution to problem (\ref{eq:surrogate prob for phi}) can then be easily obtained as
\be\label{eq:update phi}
\bm{\phi}^\star = e^{\jmath(\pi+\angle \widetilde{\mathbf{g}}_\phi)}.
\ee

\textbf{Update $\bm{\varphi}$, $\mathbf{v}$, and $\bm{\nu}$}: Using the results in (\ref{eq:AL function for phi})-(\ref{eq:kron term}) and following the same procedure as in (\ref{eq:MM for phi})-(\ref{eq:update phi}), the updates of $\bm{\varphi}$, $\mathbf{v}$, and $\bm{\nu}$ can be obtained as
\begin{subequations}\label{eq:update varphi v nu}\begin{align}
\bm{\varphi}^\star &= e^{\jmath(\pi+\angle \widetilde{\mathbf{g}}_\varphi)},\\
\mathbf{v}^\star &= e^{\jmath(\pi+\angle \widetilde{\mathbf{g}}_v)},\\
\bm{\nu}^\star &= e^{\jmath(\pi+\angle \widetilde{\mathbf{g}}_\nu)},
\end{align}\end{subequations}
where the expressions of $\widetilde{\mathbf{g}}_\varphi$, $\widetilde{\mathbf{g}}_v$, and $\widetilde{\mathbf{g}}_\nu$ can be readily obtained in a similar manner.

\textbf{Update dual variables}:
After obtaining the variables $\mathbf{r}$, $\mathbf{c}$, $\mathbf{J}$, $\mathbf{f}$, $\mathbf{w}$, $\bm{\phi}$, $\bm{\varphi}$, $\mathbf{v}$, and $\bm{\nu}$, the dual variables are updated by
\begin{subequations}\label{eq:update dual variables}\begin{align}
\zeta_i &:= \zeta_i+(\tilde{f}_i-f_i)/\rho_1,\\
\bm{\mu} &:=\bm{\mu} +  (\bm{\phi}-\bm{\varphi})/\rho_2,\\
\bm{\lambda} &:=\bm{\lambda}+ (\mathbf{v}-\bm{\phi}\otimes\bm{\varphi})/\rho_3,\\
\bm{\omega} &:=\bm{\omega}+ (\bm{\nu}-\mathbf{v})/\rho_4.
\end{align}\end{subequations}

\subsection{Summary}

\begin{algorithm}[!t]
\begin{small}
\caption{CRB-Constrained Joint Beamforming and Reflection Design}
\label{alg}
    \begin{algorithmic}[1]
    \REQUIRE $\mathbf{h}_{\text{d},\text{t}}$, $\dot{\mathbf{h}}_{\text{d},\text{t}}$, $\mathbf{h}_{\text{r},\text{t}}$, $\dot{\mathbf{h}}_{\text{r},\text{t}}$, $\mathbf{G}$, $\sigma_\text{t}^2$, $\alpha_\text{t}$, $\sigma_\text{r}^2$, $\mathbf{h}_{\text{d},k}$, $\mathbf{h}_{\text{r},k}$, $\sigma_k^2$, $\forall k$, $P_\text{t}$, $L$, $~~~~~\varepsilon$, $\rho_1$, $\rho_2$, $\rho_3$, $\rho_4$.
    \ENSURE $\mathbf{W}^\star$ and $\bm{\phi}^\star$.
        \STATE {Initialize $\bm{\phi}$ and $\mathbf{W}$ using RCG.}
        \WHILE {no convergence }
            \STATE{Update $\mathbf{J}$ and $\mathbf{f}$ by solving (\ref{eq:solve for J f}).}
            \STATE{Update $\mathbf{w}$ by solving (\ref{eq:solve for w}).}
            \STATE{Update $\bm{\phi}$ by (\ref{eq:update phi}).}
            \STATE{Update $\bm{\varphi}$, $\mathbf{v}$, and $\bm{\nu}$ by (\ref{eq:update varphi v nu}).}
            \STATE{Update dual variables $\bm{\zeta}$, $\bm{\mu}$, $\bm{\lambda}$, and $\bm{\omega}$ by (\ref{eq:update dual variables}).}
            \STATE{$\rho_1:=0.8\rho_1$, $\rho_2:=0.8\rho_2$, $\rho_3:=0.8\rho_3$, and $\rho_4:=0.8\rho_4$.}
        \ENDWHILE
        \STATE{Reshape $\mathbf{w}$ to $\mathbf{W}$.}
        \STATE{Return $\mathbf{W}^\star = \mathbf{W}$ and $\bm{\phi}^\star = \bm{\phi}$.}
    \end{algorithmic}
    \end{small}
\end{algorithm}

Given the above derivations, the proposed algorithm for CRB-constrained joint beamforming and reflection design is straightforward and summarized in Algorithm 2.
We use the same method in Sec. \ref{sec:summary and initialize} to successively initialize $\bm{\phi}$ and $\mathbf{W}$.
After that, all the variables are alternately updated until convergence.
The penalty parameters are shrunk in each iteration to accelerate the process to satisfy the equality constraints. 
For any fixed penalty parameters, the achieved objective value of problem (\ref{eq:AL function for phi}) is an upper bound of the objective value of problem (\ref{eq:subproblem with fixed W}) in solving for $\bm{\phi}$. The upper bound is gradually tightened by alternately updating each variable. 
It is noted that the objective value of (\ref{eq:AL function for phi}a) is monotonically non-increasing \cite{Sun-RSP-2017}, and the solution obtained by alternately optimization ensures convergence to a stationary point of the problem (\ref{eq:subproblem with fixed W}). 
Similarly, we see that the achieved objective value of problem (\ref{eq:AL prob for CRB}) is a lower bound of the objective value of problem (\ref{eq:opt1}), which is equivalent to the original optimization problem (\ref{eq:CRB constrained prob}), and the lower bound is tightened over iterations. 
Considering that the optimal solution to $\mathbf{r}$, $\mathbf{c}$, $\mathbf{J}$, and $\mathbf{f}$ can be obtained and the stationary points of $\mathbf{w}$ and $\bm{\phi}$ are achieved, respectively, the objective value (\ref{eq:CRB constrained prob}a) is non-decreasing.
Moreover, since the objective value (\ref{eq:CRB constrained prob}a) is upper bounded by a finite value due to the transmit power budget, any limit point of $\{\mathbf{r},\mathbf{c},\mathbf{J},\mathbf{f}, \mathbf{w}, \bm{\phi}\}$ is a stationary point of the original problem (\ref{eq:CRB constrained prob}). 
Therefore, a stationary point and a locally optimal solution can be guaranteed by Algorithm 2. 
Then, we provide a brief analysis of the computational complexity of the proposed CRB-constrained joint beamforming and reflection design algorithm. It is obvious that the update for $\mathbf{w}$ and $\mathbf{v}$ dominate the computational cost, whose complexities are of order $\mathcal{O}\{M^{3.5}(M+K)^{3.5}\}$ and $\mathcal{O}\{N^4\}$, respectively. Thus, the overall complexity of the proposed algorithm is of order $\mathcal{O}\{M^{3.5}(M+K)^{3.5}+N^4\}$.

\section{Extensions to Imperfect Self-Interference Cancellation Scenario}

In this section, we extend the proposed joint transmit beamforming and reflection design algorithms to the scenario with imperfect self-interference cancellation (SIC).
Specifically, although with the aid of advanced SIC technologies  \cite{Barneto-TMTT-2019}, \cite{Barneto-WC-2021}, there still exists residual SI that interferes with the echo signal processing of the BS receive antenna array. 
Without loss of generality, we model the residual SI signal as $\mathbf{H}_\text{SI}\mathbf{x}$, where $\mathbf{H}_\text{SI}\in\mathbb{C}^{M\times M}$ denotes the residual SI channel between transmit and receive antenna arrays \cite{He-arxiv-2023}. 
Therefore, the received signal at the BS including both target echo signal and SI plus noise can be re-formulated as  
\begin{equation}
	\mathbf{Y}_\text{r} = \alpha_\text{t}\mathbf{H}_\text{t}(\bm{\phi})\mathbf{W}\mathbf{S} + \mathbf{H}_\text{SI}\mathbf{W}\mathbf{S} + \mathbf{N}_\text{r}.\label{eq:Yr with SI}
\end{equation}
Undoubtedly, the existence of SI will influence both performance metrics and algorithm designs for the target detection and parameter estimation functions as presented below.

For the target detection function, by following the same procedure as in (\ref{eq:Yr after match filter})-(\ref{eq:SNRt}), the worst-case radar SINR that is positively proportional to the target detection probability can be expressed as 
\begin{equation}\label{eq:SINRt}\begin{aligned}
		\text{SINR}_\text{t} &= \frac{\sigma_\text{t}^2\mathbb{E}\big\{|\mathbf{u}^H(\mathbf{SS}^H\otimes\mathbf{H}_\text{t}(\bm{\phi}))\mathbf{w}|^2\big\}}
		{L\sigma_\text{r}^2\mathbf{u}^H\mathbf{u} + \mathbb{E}\big\{|\mathbf{u}^H(\mathbf{SS}^H\otimes\mathbf{H}_\text{SI})\mathbf{w}|^2\big\}} \\
		&\approx \frac{L\sigma_\text{t}^2|\mathbf{u}^H(\mathbf{I}_{K+M}\otimes\mathbf{H}_\text{t}(\bm{\phi}))\mathbf{w}|^2}{\sigma_\text{r}^2\mathbf{u}^H\mathbf{u}
			+|\mathbf{u}^H(\mathbf{I}_{K+M}\otimes \mathbf{H}_\text{SI})\mathbf{w}|^2}.
\end{aligned}\end{equation}
It is obvious that this radar SINR requirement is more complex since the SI term is associated with the transmit beamforming $\mathbf{w}$ and the receive filter $\mathbf{u}$. 
Thus, some modifications to the proposed SNR-constrained joint beamforming and reflection design in Algorithm 1 are required to handle this radar SINR constraint. 
Particularly, by using the typical Rayleigh quotient, the optimal solution $\mathbf{u}^\star$ for maximizing the worst-case radar SINR is given by
\begin{equation}\label{eq:ustar for SI}
	\mathbf{u}^\star = \frac{\mathbf{M}^{-1}(\mathbf{w})\mathbf{s}(\bm{\phi},\mathbf{w})}
	{\mathbf{s}^H(\bm{\phi},\mathbf{w})\mathbf{M}^{-1}(\mathbf{w})\mathbf{s}(\bm{\phi},\mathbf{w})},
\end{equation}
where for simplicity we respectively define the following functions with respect to $\mathbf{w}$ and $\bm{\phi}$ as 
\begin{subequations}\begin{align}
		\mathbf{M}(\mathbf{w}) &\triangleq (\mathbf{I}_{K+M}\!\otimes\! \mathbf{H}_\text{SI})\mathbf{ww}^H(\mathbf{I}_{K+M}\!\otimes\! \mathbf{H}^H_\text{SI})+\sigma_\text{r}^2\mathbf{I},\\
		\mathbf{s}(\bm{\phi},\mathbf{w}) &\triangleq (\mathbf{I}_{K+M}\otimes\mathbf{H}_\text{t}(\bm{\phi}))\mathbf{w}.
\end{align}\end{subequations}	
Then, substituting $\mathbf{u}^\star$ in (\ref{eq:ustar for SI}) into (\ref{eq:SINRt}), the worst-case radar SINR is re-formulated as 
\begin{equation}
\text{SINR}_\text{t} = {L\sigma_\text{t}^2}\mathbf{s}^H(\bm{\phi},\mathbf{w})\mathbf{M}^{-1}(\mathbf{w})\mathbf{s}(\bm{\phi},\mathbf{w}).
\end{equation}
In order to tackle the inverse term with respect to $\mathbf{w}$, we propose to utilize the idea of MM again and construct a lower-bound for $\text{SINR}_\text{t}$. 
As shown in \cite{Sun-RSP-2017}, by using the first-order Taylor expansion, a surrogate function of $\mathbf{s}^H\mathbf{M}^{-1}\mathbf{s}$ at point $(\mathbf{s}_t,\mathbf{M}_t)$ is given by 
\begin{equation}
	\mathbf{s}^H\mathbf{M}^{-1}\mathbf{s} \geq -\text{Tr}\{\mathbf{M}_t^{-1}\mathbf{s}_t\mathbf{s}_t^H\mathbf{M}_t^{-1}\mathbf{M}\}
	+ 2\Re\{\mathbf{s}_t^H\mathbf{M}_t^{-1}\mathbf{s}\},
\end{equation}
where matrix $\mathbf{M}$ must be positive-definite.
Based on this finding, a lower-bound for $\text{SINR}_\text{t}$ can be calculated as 
\begin{subequations}\begin{align}
	&\text{SINR}_\text{t}
	 \geq {L\sigma_\text{t}^2}  \big[2\Re\{\mathbf{s}^H(\bm{\phi}_t,\mathbf{w}_t)\mathbf{M}^{-1}(\mathbf{w}_t)\mathbf{s}(\bm{\phi},\mathbf{w})\} \\
	&~~ -\text{Tr}\big\{\mathbf{M}^{-1}(\mathbf{w}_t)\mathbf{s}(\bm{\phi}_t,\mathbf{w}_t)\mathbf{s}^H(\bm{\phi}_t,\mathbf{w}_t)
	\mathbf{M}^{-1}(\mathbf{w}_t)\mathbf{M}(\mathbf{w})\big\}  \big ]  \nonumber      \\
	& = {L\sigma_\text{t}^2}\big[2\Re\big\{\mathbf{g}^H_t(\mathbf{I}\!\otimes\!\mathbf{H}_\text{t}(\bm{\phi}))\mathbf{w}\big\}-|\mathbf{w}^H\mathbf{d}_t|^2
	-c_t\big],\label{eq: SINR surrogate}
\end{align}\end{subequations}
where $\mathbf{w}_t$ and $\bm{\phi}_t$ respectively represent the solution obtained in the $t$-th iteration and the variables $\mathbf{d}_t$, $\mathbf{g}_t$, and $c_t$ are defined as follows
\begin{subequations}\begin{align}
		\mathbf{d}_t&\triangleq (\mathbf{I}\otimes \mathbf{H}^H_\text{SI})\mathbf{M}^{-1}(\mathbf{w}_t)\mathbf{s}(\bm{\phi}_t,\mathbf{w}_t),\\
		\mathbf{g}_t&\triangleq\mathbf{M}^{-1}(\mathbf{w}_t) \mathbf{s}(\bm{\phi}_t,\mathbf{w}_t),\\
		c_t &\triangleq \sigma^2_\text{r}\mathbf{s}^H(\bm{\phi}_t,\mathbf{w}_t)
		\mathbf{M}^{-1}(\mathbf{w}_t)\mathbf{M}^{-1}(\mathbf{w}_t)\mathbf{s}(\bm{\phi}_t,\mathbf{w}_t).
\end{align}\end{subequations}
Now we see that the surrogate function of the radar SINR (i.e. (\ref{eq: SINR surrogate})) is concave with respect to $\mathbf{w}$, and the update for $\mathbf{w}$ can be easily solved using various existing algorithms. While given other variables, the update for $\bm{\phi}$, $\mathbf{r}$ and $\mathbf{c}$ can be re-arranged as similar forms as those in (\ref{eq:solve phi org}), (\ref{eq:update rk}), and (\ref{eq:update ck}), respectively. Thus, Algorithm 1 can be utilized to solve the SINR-constrained joint design problem under an imperfect SIC scenario.

For the parameter estimation function, we first vectorize the SI-corrupted received signal $\mathbf{Y}_\text{r}$ in (\ref{eq:Yr with SI}) as 
\begin{equation}
	\mathbf{y}_\text{r} = \alpha_\text{t}\text{vec}\{\mathbf{H}_\text{t}(\bm{\phi})
	\mathbf{WS}\} + \text{vec}\{\mathbf{H}_\text{SI}\mathbf{WS}\} + \mathbf{n}_\text{r},
\end{equation}
from which we have $\mathbf{y}_\text{r}\sim\mathcal{CN}(\bm{\eta},\mathbf{R}_\text{r})$ with $\mathbf{R}_\text{r} = \sigma^2_\text{r}\mathbf{I}_{ML}+\mathbf{I}_L\otimes \mathbf{H}_\text{SI}\mathbf{WW}^H\mathbf{H}_\text{SI}^H$. 
As presented in \cite{estimation theory}, the FIM under the imperfect SIC scenario is given by 
\be
\mathbf{F}_\text{IM}(i,j) = 2\Re\Big\{\frac{\partial^H\bm{\eta}}
{\partial\xi_i}\mathbf{R}_\text{r}^{-1}\frac{\partial\bm{\eta}}
{\partial\xi_j}\Big\}.
\label{eq:define FIM SI}\ee
Following the same procedure as that in (\ref{eq:derivatives for eta}) and (\ref{eq:each ele of FIM}), the elements of $\mathbf{F}_\text{IM}$ can be respectively calculated, e.g., the element of the first row and the second column is expressed as 
\begin{equation}
	F_{\theta_1,\theta_2}= \frac{2L|\alpha_\text{t}|^2}{\sigma_\text{r}^2}\Re\{\text{Tr}\{
	\ddot{\mathbf{H}}_\text{t}(\bm{\phi})\mathbf{WW}^H\dot{\mathbf{H}}^H_\text{t}(\bm{\phi})\mathbf{R}^{-1}(\mathbf{W})\}\},
\end{equation}
where we define the function with respect to $\mathbf{W}$ as $\mathbf{R}(\mathbf{W}) \triangleq \mathbf{I}_{M}+\mathbf{H}_\text{SI}\mathbf{WW}^H\mathbf{H}_\text{SI}^H/\sigma^2_\text{r}$.
We observe that the residual SI introduces the term $\mathbf{H}_\text{SI}\mathbf{WW}^H\mathbf{H}_\text{SI}^H$ to FIM, which will lead to very complicated optimization with respect to $\mathbf{W}$. 
In order to simplify the optimization on $\mathbf{W}$, in each iteration we update $\mathbf{W}$ with fixed $\mathbf{R}(\mathbf{W})$, and then calculate $\mathbf{R}(\mathbf{W})$ using the resulting $\mathbf{W}$. In the sequel, Algorithm 2 can be utilized to solve the CRB-constrained joint design problem under an imperfect SIC scenario.

\section{Simulation Results}\label{sec:simulation results}

In this section, we present simulation results to verify the advantages of the proposed SNR-constrained and CRB-constrained joint beamforming and reflection designs.
We assume that  $M=6$, $K = 4$, $\sigma_\text{r}^2 = \sigma_k^2 = -90$dBm, $\forall k$, $\sigma_\text{t}^2 = 1$, and $L = 1024$.
We set the distances of the BS-RIS, the RIS-target, and the RIS-user links as $d_\text{BR} = 50$m, $d_\text{RT} = 3$m, and $d_\text{RU}=8$m, respectively.
Furthermore we set $\theta_\text{BR} = \theta_2 = \frac{\pi}{4}$, $\theta_\text{RB} = -\frac{\pi}{4}$, and $\theta_1$ is calculated as $\theta_1 = \text{atan}(\frac{d_\text{BR}\sin\theta_\text{BR}-d_\text{RT}\cos\theta_2 }
{d_\text{BR}\cos\theta_\text{BR}+d_\text{RT}\sin\theta_2})$.
The distances of the BS-target and BS-user can then be calculated accordingly.
We adopt a typical distance-dependent path-loss model \cite{Wu-TWC-2019} and set the path-loss exponents for the BS-RIS, RIS-target, RIS-user, BS-target and BS-user links as 2.2, 2.2, 2.3, 2.4, and 3.5, respectively.
Since the users are several meters farther away from the target, the reflected signals from the target to the users are ignored due to severe channel fading.
In addition, the Rician factor for the BS-RIS/user and RIS-user links is set as $\kappa = 3$dB. 
The residual SI channel is modeled as $\mathbf{H}_\text{SI}(i,j) = \sqrt{\alpha_\text{SI}}e^{-\jmath 2\pi d_{i,j}/\lambda}$, where $\alpha_\text{SI}$, $d_{i,j}$ and $\lambda$ denote the power of residual SI, the distance between the $i$-th transmit antenna and the $j$-th receive antenna, and the wavelength, respectively. For simplicity, we set $\alpha_\text{SI} = -110$dB and let $e^{-\jmath 2\pi d_{i,j}/\lambda}$ be a unit-modulus variable with random phase \cite{He-arxiv-2023}.
Since our focus is target DoA estimation, we assume $\alpha_\text{t}=1$ for simplicity.

\begin{figure}[!t]
  \centering
  \includegraphics[width=\linewidth]{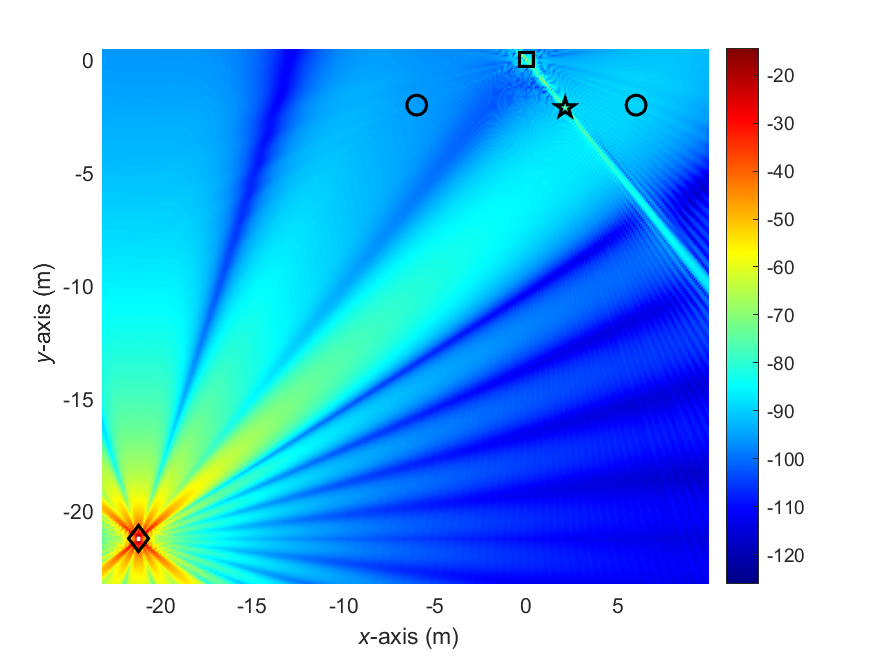}\\
  \caption{Enhanced beampattern of the RIS-assisted system (BS: diamond; RIS: square; target: star; users: circles).}\label{fig:beampattern}
\end{figure}
\subsection{Illustration of Radar Sensing Performance}

We first visually demonstrate the communications and radar sensing functions by plotting the enhanced beampattern of the RIS-assisted ISAC system in Fig. \ref{fig:beampattern}.
We clearly see that the BS generates strong beams towards the areas where the RIS, the target, and the users are located, meanwhile the RIS forms multiple passive beams to direct the signals towards the target and the users.

\subsection{Convergence Performance of the Proposed Algorithms}
\begin{figure}[t]
\centering
\subfigure[Convergence of Algorithm 1.]{
\begin{minipage}{0.225\textwidth}
\centering
\includegraphics[width = \linewidth]{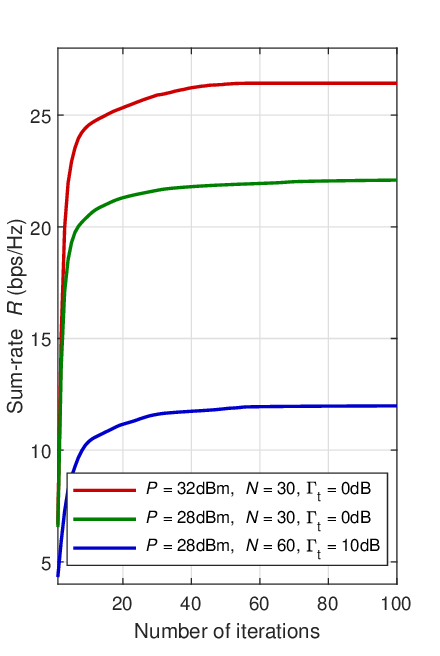}\label{fig:convergence1}
\end{minipage}
}
\subfigure[Convergence of Algorithm 2.]{
\begin{minipage}{0.225\textwidth}
\centering
\includegraphics[width = \linewidth]{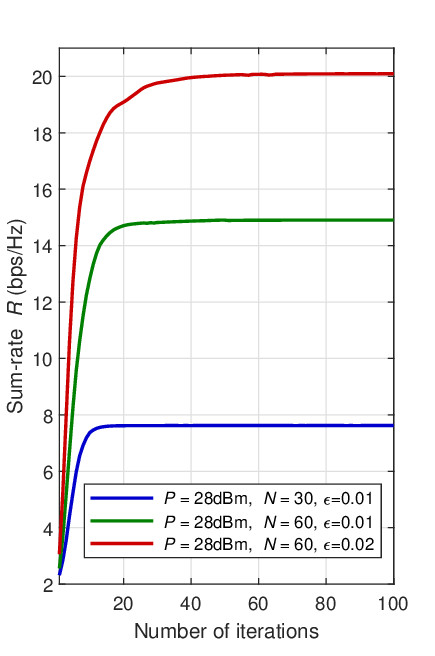}\label{fig:convergence2}
\end{minipage}
}
\caption{Convergence performance of the proposed algorithms.}\label{fig:convergence}
\end{figure}

We show the convergence performance of Algorithm 1 and Algorithm 2 in Figs. \ref{fig:convergence1} and \ref{fig:convergence2}, respectively.
The achievable sum-rate versus the number of iterations under different settings is presented to verify the convergence of our proposed algorithms.
We observe that both Algorithm 1 and Algorithm 2 converge within 100 iterations.
Moreover, only small increases in the achievable sum-rate can be acquired after 30 iterations, which verifies the fast convergence and the effectiveness of the proposed algorithms.

\subsection{Impact of Transmit Power}

To verify the advantages of the proposed joint beamforming and reflection designs (denoted as \textbf{``Proposed''}), we also include the following schemes for comparisons.
\begin{itemize}
  \item \textbf{``Comm only''}: Only the multi-user communication function is optimized for the considered system. The proposed joint beamforming and reflection design algorithm is utilized without considering the radar sensing constraint.
  \item \textbf{``BF only''}: With a fixed $\bm{\phi}$ which is determined by maximizing the sum channel gain, the beamformers $\mathbf{W}$ and $\mathbf{u}$ are iteratively solved for using Algorithm 1 for the SNR-constrained scenario, or $\mathbf{W}$ is iteratively solved for using Algorithm 2 for the CRB-constrained scenario.
  \item \textbf{``Separate''}: The reflection coefficients $\bm{\phi}$, the radar beamformer $\mathbf{W}_\text{r}$, the communication beamformer $\mathbf{W}_\text{c}$, and the receive filter $\mathbf{u}$ are separately optimized. In particular, $\bm{\phi}$ is obtained by maximizing the sum channel gain, $\mathbf{W}_\text{r}$ is determined by solving the power minimization problem under the radar SNR or CRB constraint, $\mathbf{W}_\text{c}$ is then designed by solving the sum-rate maximization problem under the power constraint, and finally for the SNR-constrained scenario $\mathbf{u}$ is given by (\ref{eq:update u}).
\end{itemize}
In simulations, we consider both perfect SIC and imperfect SIC scenarios, which are represented by solid and dashed lines, respectively. To distinguish the imperfect SIC scenario, we also mark it with \textbf{``w/ SI''}.

\begin{figure}[!t]
\centering
\subfigure[SNR-constrained ($\Gamma_\text{t}=7$dB).]{
\begin{minipage}{0.225\textwidth}
\centering
\includegraphics[width = \linewidth]{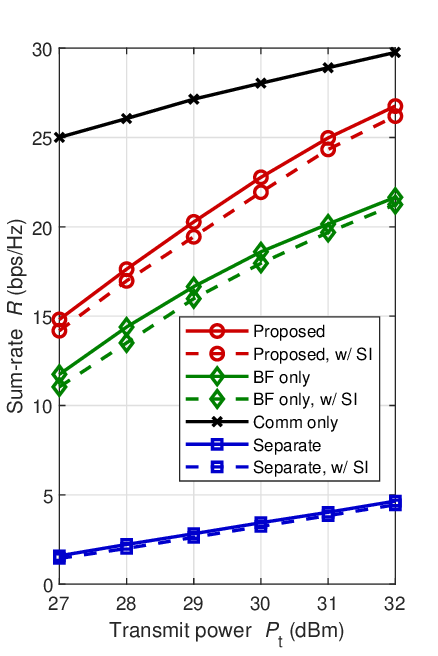}\label{fig:SR vs P}
\end{minipage} }
\subfigure[CRB-constrained ($\varepsilon=0.02$).]{
\begin{minipage}{0.225\textwidth}
\centering
\includegraphics[width = \linewidth]{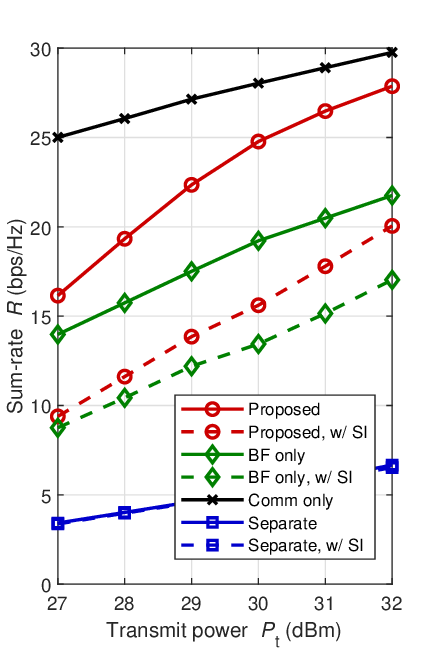}\label{fig:SR vs P for CRB}
\end{minipage} }
\caption{Sum-rate versus transmit power $P_\text{t}$ ($N = 49$).}\label{fig:SR vs power}
\end{figure}

The achievable sum-rate versus transmit power $P_\text{t}$ for the SNR-constrained joint beamforming and reflection design is presented in Fig. \ref{fig:SR vs P}.
We observe that the schemes that jointly design the radar and communication beamforming achieve a remarkable performance improvement compared with the ``Separate'' scheme.
Undoubtedly, the ``Comm only'' scheme achieves the best sum-rate performance, and the proposed scheme is much better than the ``BF only'' scheme.
Moreover, we observe that the performance gap between ``Comm only'' and the proposed scheme becomes smaller with the increase of $P_\text{t}$, since more transmit power is exploited to improve the communication performance for a fixed radar sensing requirement. 
Besides, it is clear that residual SI causes certain performance losses to all ISAC schemes.
The performance of the CRB-constrained design is shown in Fig. \ref{fig:SR vs P for CRB}, in which the same performance relationship as that for the SNR-constrained scenario is observed and similar conclusions can be drawn. 
Moreover, compared to Fig. \ref{fig:SR vs P} we notice that the residual SI has a more pronounced effect on the parameter estimation performance. This is because the target detection performance depends only on the power of interference and useful signals, while the parameter estimation performance is also sensitive to the phase of the received signals.

\subsection{Impact of the Number of Reflecting Elements}

\begin{figure}[!t]
\centering
\subfigure[SNR-constrained ($\Gamma_\text{t} = 7$dB).]{
\begin{minipage}{0.225\textwidth}
\centering
\includegraphics[width = \linewidth]{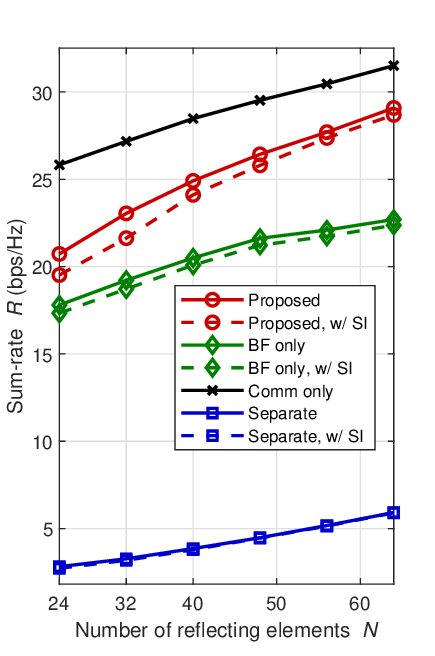}\label{fig:SR vs N}
\end{minipage} }
\subfigure[CRB-constrained ($\varepsilon=0.02$).]{
\begin{minipage}{0.225\textwidth}
\centering
\includegraphics[width = \linewidth]{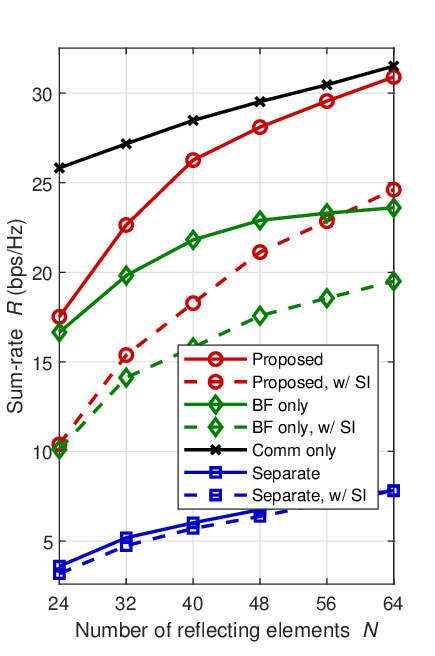}\label{fig:SR vs N for CRB}
\end{minipage} }
\caption{Sum-rate versus the number of reflecting elements $N$ ($P = 32$dBm).}\label{fig:SR vs value N}
\end{figure}
\begin{figure}[!t]
\centering
\subfigure[Sum-rate versus SNR $\Gamma_\text{t}$.]{
\begin{minipage}{0.225\textwidth}
\centering
\includegraphics[width = \linewidth]{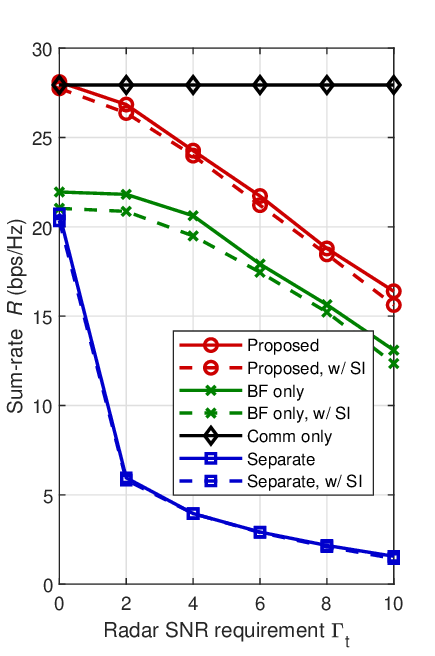}\label{fig:SR vs gammat}
\end{minipage}
}
\subfigure[Sum-rate versus CRB $\varepsilon$.]{
\begin{minipage}{0.225\textwidth}
\centering
\includegraphics[width = \linewidth]{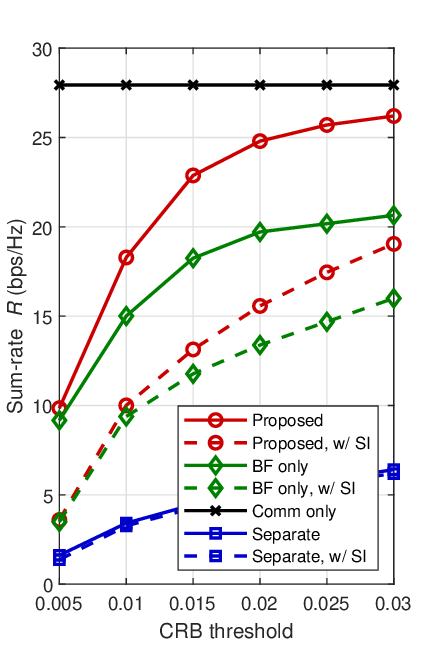}\label{fig:SR vs CRB}
\end{minipage}
}
\caption{Impact of radar sensing performance ($N = 49$, $P = 30$dBm).}\label{fig:SR vs radar}
\end{figure}

Next, we illustrate the achievable sum-rate versus the number of reflecting elements $N$ in Fig. \ref{fig:SR vs value N}.
It is obvious that more reflecting elements provide larger passive beamforming gains since they can exploit more DoFs to manipulate the propagation environment.
Moreover, comparing Fig. \ref{fig:SR vs N} and Fig. \ref{fig:SR vs N for CRB}, we see that additional reflecting elements provide more pronounced performance gains for the CRB-constrained schemes.
Specifically, the achievable sum-rate increases by 40\% in the target detection scenario, and 76\% in the target DoA estimation scenario.
This phenomenon implies that a larger RIS not only directs more energy to the target for detection but more importantly, improves the accuracy of the DoA estimation.

\subsection{Impact of the Radar Sensing Performance}

The impact of different radar sensing requirements is shown in Fig. \ref{fig:SR vs radar}.
We clearly observe the performance trade-off between multi-user communications and radar target detection or DoA estimation.
In addition, a huge performance degradation for the ``Separate'' scheme occurs when the radar SNR changes from 0dB to 2dB.
This is because a tighter target detection constraint requires more power in designing $\mathbf{W}_\text{r}$,  thus leaving less power for optimizing $\mathbf{W}_\text{c}$ to maximize the sum-rate.

\section{Conclusions}\label{sec:conclusions}

In this paper, we considered a general case for RIS-assisted ISAC systems, in which both the direct and reflected links contribute to MU-MISO communications and radar sensing.
In addition to the sum-rate performance metric for multi-user communications, we derived the radar SNR metric for target detection and the CRB for estimating target DoAs.
We formulated the optimization problems that maximize the sum-rate as well as satisfy a worst-case radar SNR or CRB constraint, the transmit power budget, and the unit-modulus constraint of the RIS reflection coefficients.
Efficient alternating algorithms were developed to solve the resulting non-convex problems.
Simulation results verified the advantages of the proposed algorithms, illustrated the performance improvements introduced by more resources, and demonstrated the performance trade-off between communications and radar sensing with limited resources.
Motivated by this work, we will further explore more application scenarios of RIS in ISAC systems, e.g., active/multiple RIS, and some practical issues including clutter suppression.

\setcounter{TempEqCnt}{\value{equation}}
\setcounter{equation}{74}
\begin{figure*}[t]
\begin{subequations}\label{eq:ftilde}\begin{align}
\mathcal{F}_1(\mathbf{W}, \bm{\phi}) &\triangleq \text{Tr}\{\dot{\mathbf{H}}_\text{t}(\bm{\phi})\mathbf{WW}^H\dot{\mathbf{H}}^H_\text{t}(\bm{\phi})\} =
\sum_{k=1}^{K+M}\mathbf{w}_k^H\underbrace{\dot{\mathbf{H}}^H_\text{t}(\bm{\phi})\dot{\mathbf{H}}_\text{t}(\bm{\phi})}_{\mathbf{A}_1}\mathbf{w}_k
= \sum_{k=1}^{K+M}(\mathbf{q}_{1,k}+\mathbf{U}_{1,k}\bm{\phi})^H(\mathbf{q}_{1,k}+\mathbf{U}_{1,k}\bm{\phi})\\
&= \underbrace{\sum_{k=1}^{K+M}\mathbf{q}_{1,k}^H\mathbf{q}_{1,k}}_{q_1} + 2\Re\Big\{\bm{\phi}^H\underbrace{\sum_{k=1}^{K+M}\mathbf{U}_{1,k}^H\mathbf{q}_{1,k}}_{\mathbf{u}_1}\Big\} +\bm{\phi}^H\underbrace{\sum_{k=1}^{K+M}\mathbf{U}_{1,k}^H\mathbf{U}_{1,k}}_{\mathbf{U}_1}\bm{\phi},\\
\mathcal{F}_2(\mathbf{W}, \bm{\phi}) &\triangleq\text{Tr}\{\ddot{\mathbf{H}}_\text{t}(\bm{\phi})\mathbf{WW}^H\dot{\mathbf{H}}^H_\text{t}(\bm{\phi})\} =
\text{Tr}\{\mathbf{A}_2\mathbf{WW}^H\} = \overline{\mathbf{u}}_2^H\bm{\phi} + \bm{\phi}^H\mathbf{U}_2\bm{\phi} + \overline{\mathbf{z}}_2^H(\bm{\phi}\!\otimes\!\bm{\phi}) +  \bm{\phi}^H\mathbf{C}_2(\bm{\phi}\!\otimes\!\bm{\phi}),   \\
\mathcal{F}_3(\mathbf{W}, \bm{\phi}) &\triangleq\text{Tr}\{\mathbf{H}_\text{t}(\bm{\phi})\mathbf{WW}^H\dot{\mathbf{H}}^H_\text{t}(\bm{\phi})\} =
\text{Tr}\{\mathbf{A}_3\mathbf{WW}^H\} = q_3 + \bm{\phi}^H\mathbf{u}_3 +\overline{\mathbf{u}}_3^H\bm{\phi} +
\bm{\phi}^H\mathbf{U}_3\bm{\phi} +\overline{\mathbf{z}}_3^H(\bm{\phi}\!\otimes\!\bm{\phi}) + \bm{\phi}^H\mathbf{C}_3(\bm{\phi}\!\otimes\!\bm{\phi}),\\
\mathcal{F}_4(\mathbf{W}, \bm{\phi}) &\triangleq\text{Tr}\{\ddot{\mathbf{H}}_\text{t}(\bm{\phi})\mathbf{WW}^H\ddot{\mathbf{H}}^H_\text{t}(\bm{\phi})\} = \text{Tr}\{\mathbf{A}_4\mathbf{WW}^H\} = \bm{\phi}^H\mathbf{U}_4\bm{\phi} +
2\Re\big\{(\bm{\phi}\!\otimes\!\bm{\phi})^H\overline{\mathbf{C}}_4\bm{\phi}\big\} + (\bm{\phi}\!\otimes\!\bm{\phi})^H\mathbf{Z}_4(\bm{\phi}\!\otimes\!\bm{\phi}), \\
\mathcal{F}_5(\mathbf{W}, \bm{\phi}) &\triangleq\text{Tr}\{\mathbf{H}_\text{t}(\bm{\phi})\mathbf{WW}^H\ddot{\mathbf{H}}^H_\text{t}(\bm{\phi})\} = \text{Tr}\{\mathbf{A}_5\mathbf{WW}^H\} \non \\
& = \bm{\phi}^H\mathbf{u}_5 +
\bm{\phi}^H\mathbf{U}_5\bm{\phi} + (\bm{\phi}\!\otimes\!\bm{\phi})^H\mathbf{z}_5 + (\bm{\phi}\!\otimes\!\bm{\phi})^H\mathbf{Z}_5(\bm{\phi}\!\otimes\!\bm{\phi})+ (\bm{\phi}\!\otimes\!\bm{\phi})^H\overline{\mathbf{C}}_5\bm{\phi} + \bm{\phi}^H\mathbf{C}_5(\bm{\phi}\!\otimes\!\bm{\phi}),\\
\mathcal{F}_6(\mathbf{W}, \bm{\phi}) &\triangleq\text{Tr}\{\mathbf{H}_\text{t}(\bm{\phi})\mathbf{WW}^H\mathbf{H}^H_\text{t}(\bm{\phi})\} = \text{Tr}\{\mathbf{A}_6\mathbf{WW}^H\} \non\\
& = q_6 + 2\Re\big\{\bm{\phi}^H\mathbf{u}_6\big\} +\bm{\phi}^H\mathbf{U}_6\bm{\phi}+ 2\Re\big\{(\bm{\phi}\!\otimes\!\bm{\phi})^H\mathbf{z}_6\big\} + (\bm{\phi}\!\otimes\!\bm{\phi})^H\mathbf{Z}_6(\bm{\phi}\!\otimes\!\bm{\phi}) + 2\Re\big\{(\bm{\phi}\!\otimes\!\bm{\phi})^H\overline{\mathbf{C}}_6\bm{\phi}\big\}.
\end{align}
\end{subequations}
\rule[-0pt]{18.5 cm}{0.05em}
\end{figure*}
\setcounter{equation}{\value{TempEqCnt}}

\begin{appendices}
\section{}\setcounter{equation}{75}

According to (\ref{eq:vec Yr}), the derivatives of $\bm{\eta}$ with respect to each parameter can be calculated as
\begin{subequations}\label{eq:derivatives for eta}
\begin{align}
\frac{\partial\bm{\eta}}{\partial\theta_1} &= \alpha_\text{t}\text{vec}\{\dot{\mathbf{H}}_\text{t}(\bm{\phi})\mathbf{WS}\},\\
\frac{\partial\bm{\eta}}{\partial\theta_2} &= \alpha_\text{t}\text{vec}\{\ddot{\mathbf{H}}_\text{t}(\bm{\phi})\mathbf{WS}\},\\
\frac{\partial\bm{\eta}}{\partial\bm{\alpha}} &= [1~\jmath]^T \otimes\text{vec}\{{\mathbf{H}}_\text{t}(\bm{\phi})\mathbf{WS}\},
\end{align}
\end{subequations}
where $\dot{\mathbf{H}}_\text{t}(\bm{\phi})$ and $\ddot{\mathbf{H}}_\text{t}(\bm{\phi})$ denote the partial derivatives of $\mathbf{H}_\text{t}(\bm{\phi})$ with respect to $\theta_1$ and $\theta_2$, respectively.
Thus, plugging (\ref{eq:derivatives for eta}) into (\ref{eq:define FIM}), the elements of $\mathbf{F}_\text{IM}$ can be calculated as
\begin{subequations}\label{eq:each ele of FIM}
\begin{align}
F_{\theta_1,\theta_1} &= \frac{2|\alpha_\text{t}|^2}{\sigma_\text{r}^2}\Re\{\text{vec}^H\{\dot{\mathbf{H}}_\text{t}(\bm{\phi})\mathbf{WS}\}
\text{vec}\{\dot{\mathbf{H}}_\text{t}(\bm{\phi})\mathbf{WS}\}\}\non\\
& = \frac{2L|\alpha_\text{t}|^2}{\sigma_\text{r}^2}\text{Tr}\{\dot{\mathbf{H}}_\text{t}(\bm{\phi})
\mathbf{WW}^H\dot{\mathbf{H}}^H_\text{t}(\bm{\phi})\}, \\
F_{\theta_1,\theta_2} &= \frac{2L|\alpha_\text{t}|^2}{\sigma_\text{r}^2}\Re\{\text{Tr}\{
\ddot{\mathbf{H}}_\text{t}(\bm{\phi})\mathbf{WW}^H\dot{\mathbf{H}}^H_\text{t}(\bm{\phi})\}\},\\
F_{\theta_2,\theta_1} &= F_{\theta_1,\theta_2},\\
F_{\theta_2,\theta_2} & = \frac{2L|\alpha_\text{t}|^2}{\sigma_\text{r}^2}\text{Tr}\{
\ddot{\mathbf{H}}_\text{t}(\bm{\phi})\mathbf{WW}^H\ddot{\mathbf{H}}^H_\text{t}(\bm{\phi})\}, \\
\mathbf{F}_{\theta_1,\bm{\alpha}^T} & =  \frac{2L}{\sigma_\text{r}^2}\Re\{\text{Tr}\{\alpha^*_\text{t}
\mathbf{H}_\text{t}(\bm{\phi})\mathbf{WW}^H\dot{\mathbf{H}}^H_\text{t}(\bm{\phi})\}[1~~\jmath]\},  \\
\mathbf{F}_{\theta_2,\bm{\alpha}^T} & = \frac{2L}{\sigma_\text{r}^2}\Re\{\text{Tr}\{\alpha^*_\text{t}
\mathbf{H}_\text{t}(\bm{\phi})\mathbf{WW}^H\ddot{\mathbf{H}}^H_\text{t}(\bm{\phi})\}[1~~\jmath]\},\\
\mathbf{F}_{\bm{\alpha},\bm{\alpha}^T} & = \frac{2L}{\sigma_\text{r}^2}\text{Tr}\{\mathbf{H}_\text{t}(\bm{\phi})
\mathbf{WW}^H\mathbf{H}^H_\text{t}(\bm{\phi})\}\mathbf{I}_2.
\end{align}\end{subequations}
It is noted that we assume that $\mathbf{SS}^H = L\mathbf{I}_K$ due to the fact that sufficient samples are usually collected for parameter estimation.
Based on the above derivations, the sub-matrices of $\mathbf{F}$ can be constructed as
\be
\mathbf{F}_{\bm{\theta}\bm{\theta}^T} = \left[\begin{array}{cc}
F_{\theta_1,\theta_1} & F_{\theta_1,\theta_2}\\
F_{\theta_2,\theta_1} & F_{\theta_2,\theta_2}\end{array}\right],~~
\mathbf{F}_{\bm{\theta}\bm{\alpha}^T} = \left[\begin{array}{c}
F_{\theta_1,\bm{\alpha}^T}\\
F_{\theta_2,\bm{\alpha}^T}\end{array}\right].
\ee

\section{}

In order to re-arrange each element of $\mathbf{F}_\text{IM}$ as explicit expressions with respect to $\mathbf{W}$ and $\bm{\phi}$, we first utilize the transformation $\text{vec}\{\mathbf{ABC}\} = (\mathbf{C}^T\otimes\mathbf{A})\text{vec}\{\mathbf{B}\}$ to re-formulate the term $\mathbf{H}_\text{t}(\bm{\phi})\mathbf{w}_k$ as
\be\label{eq:Hwk convert}
\begin{aligned}
 &\mathbf{H}_\text{t}(\bm{\phi})\mathbf{w}_k = (\underbrace{\mathbf{w}_k^T\mathbf{h}_\text{d,t}\mathbf{G}^T\!\text{diag}\{\!\mathbf{h}_\text{r,t}\!\} \!+\!
\mathbf{w}_k^T\mathbf{G}^T\!\text{diag}\{\!\mathbf{h}_\text{r,t}\!\}\!\otimes\!\mathbf{h}_\text{d,t}}_{\mathbf{U}_{0,k}})\bm{\phi}\\
&+\!
(\underbrace{\mathbf{w}_k^T\mathbf{G}^T\!\text{diag}\{\mathbf{h}_\text{r,t}\}\!\otimes\!\mathbf{G}^T\!\text{diag}\{\mathbf{h}_\text{r,t}\}}_{\mathbf{Z}_{0,k}})
\text{vec}\{\bm{\phi\phi}^T\!\}\!+\!\underbrace{\mathbf{h}_\text{d,t}\mathbf{h}_\text{d,t}^T\mathbf{w}_k}_{\mathbf{q}_{0,k}}.\\
\end{aligned}\ee
Similarly, the terms $\dot{\mathbf{H}}_\text{t}(\bm{\phi})\mathbf{w}_k$ and $\ddot{\mathbf{H}}_\text{t}(\bm{\phi})\mathbf{w}_k$ can be re-formulated as
\begin{subequations}\label{eq:Hwk convert2}\begin{align}
\dot{\mathbf{H}}_\text{t}(\bm{\phi})\mathbf{w}_k &= \mathbf{q}_{1,k} +\mathbf{U}_{1,k}\bm{\phi}, \\
\ddot{\mathbf{H}}_\text{t}(\bm{\phi})\mathbf{w}_k &= \mathbf{U}_{2,k}\bm{\phi} + \mathbf{Z}_{2,k}\text{vec}\{\bm{\phi\phi}^T\}.
\end{align}\end{subequations}
Then, the functions $\mathcal{F}_1(\mathbf{W}, \bm{\phi})\sim\mathcal{F}_6(\mathbf{W}, \bm{\phi})$ are defined and re-arranged in (\ref{eq:ftilde}) presented at the top of the next page.
Considering that the derivations are similar and straightforward, we only present the detailed derivations and expressions for $\mathcal{F}_1(\mathbf{W}, \bm{\phi})$.
The details for other variables irrelevant to $\bm{\phi}$ can be obtained in the same way and are omitted in this paper due to space limitations.

\end{appendices}

\end{document}